\documentclass[aps, prl, twocolumn, showpacs, superscriptaddress, 10pt, footinbib, floatfix, longbibliography]{revtex4-1}
\usepackage{graphicx}  
\usepackage{xcolor}
\usepackage{amsmath, amssymb}   
\usepackage{amsfonts}
\usepackage{microtype}
\usepackage{dsfont}

\usepackage{dsfont}
\usepackage{hyperref}
\usepackage{bm}

\usepackage[boxed,vlined,figure]{algorithm2e}

\begin{document}

\title{State Aggregations in Markov Chains and Block Models of Networks}

\author{Mauro Faccin}
\affiliation{ICTEAM, Universit\'e catholique de Louvain, Belgium}
\author{Michael T. Schaub}
\affiliation{Department of Engineering Science, University of Oxford, UK}
\affiliation{Department of Computer Science, RWTH Aachen University, Germany}
\author{Jean-Charles Delvenne}
\affiliation{ICTEAM, Universit\'e catholique de Louvain, Belgium}
\affiliation{CORE, Universit\'e catholique de Louvain, Belgium}

\begin{abstract}
  We consider state-aggregation schemes for Markov chains from an information-theoretic perspective.
  Specifically, we consider aggregating the states of a Markov chain such that the mutual information of the aggregated states separated by $T$ time steps is maximized.
  We show that for $T=1$ this recovers the maximum-likelihood estimator of the degree-corrected stochastic block model as a particular case, which enables us to explain certain features of the likelihood landscape of this generative network model from a dynamical lens.
  We further highlight how we can uncover coherent, long-range dynamical modules for which considering a timescale $T\gg 1$ is essential.
  We demonstrate our results using synthetic flows and real-world ocean currents, where we are able to recover the fundamental features of the surface currents of the oceans.
\end{abstract}


\maketitle

Systems comprising the interactions of many entities often exhibit complex dynamics that unfold within a large state space.
A powerful idea to tame this complexity is to project the system state $x_t$ at each time $t$ onto a significantly smaller space, and replace the original dynamics, say of the form  ${x_{t+1} = f(x_t,x_{t-1},\ldots)}$, with the simpler dynamics $y_{t+1} = g(y_t,y_{t-1},\ldots)$ of the projected state $y_t$.
Such techniques abound in physics and other fields under headings such as model order reduction, coarse graining, variable or state aggregation, mode decomposition, or dimensionality reduction~\cite{simon1961aggregation,moore1981principal,juang1985eigensystem,meyer1989stochastic,Crutchfield1989,coifman2006diffusion,noid2008multiscale,schilders2008model,Rosvall29012008,Delvenne20072010,kutz2016dynamic}.

The success of these methods hinges on the choice of a projection $y_t=h(x_t)$ that retains the salient features of the original dynamics.
For example, for a linear dynamics, a small subspace spanned by its dominant, low-frequency eigenmodes governs the long-term behavior.
The neglected eigenmodes correspond to high-frequency modes describing short-lived transients.
Projecting $x_t$ onto the slow eigenmodes yields a system description $y_t$ with theoretical guarantees on the reconstruction error of the original dynamics~\cite{schilders2008model,Benner2017,kutz2016dynamic}.
Accordingly, spectral techniques such as generalized Perron cluster cluster analysis (GenPCCA)~\cite{fackeldey2018spectral}, which extract the dominant subspaces of a dynamics, have been proposed to address the problem of state aggregation.
In other situations, we may also prefer to extract nondominant eigenvectors corresponding to medium or fast timescales~\cite{coderch1983hierarchical,Monshizadeh2014,Schaub2016}.

Here we consider a stationary Markov process on a discrete state space $\mathcal X$ and explore information-theoretic strategies to find state aggregations, that are akin to a nonlinear version of choosing between the slow and fast frequency modes.
Given a state-aggregation $y_t=h(x_t)$, we study the time-lagged mutual information $\mathcal I_T$ between the new state variables $y_t$ and $y_{t+T}$ for any timescale $T$.
We call $\mathcal I_T$ the autoinformation of the state aggregation scheme.
Related information-theoretic ideas include influential works such as the information bottleneck method~\cite{Tishby2000}, approaches from computational mechanics~\cite{shalizi2001computational}, or the map equation~\cite{Rosvall29012008} (see the Supplemental Material (SM) for a discussion of related methods).

We demonstrate that our approach offers a fresh perspective on the problem of state aggregation.
Specifically, we show that maximizing the autoinformation for unit timescales ($T=1$) is under certain conditions equivalent to maximizing the likelihood of a degree-corrected stochastic block model (DCSBM)~\cite{Dasgupta2004,karrer2011dcsbm}, a popular technique to recover community structure in networks~\cite{fortunato2010community,fortunato2016community,schaub2017many}.
Leveraging our dynamical perspective we can thus pinpoint problems inherent to assumptions underlying the DCSBM.
We further show how the time parameter $T$ of the autoinformation leads to a nonlinear transformation mitigating these problems.
Our scheme is thus particularly relevant for the analysis of trajectory data with trends emerging over longer timescales, which we illustrate by analyzing an ocean drifter dataset, where we can reveal dominant patterns such as ocean currents over long timescales.

\paragraph{Autoinformation between aggregated states.}
Consider a state aggregation $y_t=h(x_t)$ that maps the discrete state $x_t\in \mathcal X$ from a space of cardinality  $|\mathcal X|\!=\!N$ onto a new state $y_t\in \mathcal Y$ in a smaller space of size $|\mathcal Y|\!=K\! \le\! N$.
This induces a partition of $\mathcal X$ into \emph{aggregation classes}: sets of states in $\mathcal X$ mapped to the same aggregated state in $\mathcal Y$.
Applying the mapping $h$ to each observed state $x_t$ of the original trajectory yields a new trajectory that can be described by a stochastic dynamical system $y_{t+1} = g(y_t,y_{[t-1:-\infty]})$.
Here the symbol $y_{[\tau_1,\tau_2]}$ denotes the sequence of states $y_{\tau_1},\ldots,y_{\tau_2}$ from $\tau_1$ until $\tau_2$.

To find an aggregation $y_t = h(x_t)$ whose states are informative about the evolution of the dynamics at the next time step, we seek a mapping $h$ for which the mutual information $I(y_{t+1},y_t)$ is as high as possible.
It involves two terms of opposite signs:
\begin{equation}\label{eq:mutual_info}
  I(y_{t+1},y_t) = I(y_{t+1};y_{[t,-\infty]})  - I(y_{t+1}; y_{[t-1,-\infty]} | y_t).
\end{equation}
Maximizing $I(y_{t+1};y_{[t,-\infty]})$ favors state aggregations that are as deterministic (or predictable) as possible.
Minimizing $I(y_{t+1}; y_{[t-1,-\infty]}| y_t)$, however, leads to aggregations that are as Markovian as possible.
Indeed, this term quantifies how much $y_t$ deviates from a Markov process~\cite{faccin2017entrograms}: it is zero for a Markov process and positive otherwise (note that even if $x_t$ is a Markov process, the aggregated system $y_t = h(x_t)$ is not Markov in general; it is Markov if and only if the aggregation classes form a so-called lumpable partition of the transition matrix, see the SM).

We view Eq.~\eqref{eq:mutual_info} as a nonlinear counterpart to the unit time-lag linear autocorrelation of real-valued time series, which is pivotal for analyzing observables of linear dynamical systems, e.g., in signal processing or in the context of the fluctuation-dissipation theorem.
Therefore, we call $I(y_{t+1};y_t)$ the one-step autoinformation of the aggregated process.
By the same rationale we define the ($T$-step) autoinformation of the state aggregation $h$ as:
\begin{subequations}
\begin{align}\label{eq:autoinfo}
    \mathcal{I}_{T}(h) &:=I(h(x_{t+T}); h(x_{t}))=I(y_{t+T}; y_{t}) \\
                       &\;= H(y_{t}) - H(y_{t+T} | y_{t}),\label{eq:autoinfo-bis}
\end{align}
\end{subequations}
where $H(y_{t})=H(y_{t+T})$ is the Shannon entropy of the aggregated state variables.
Writing the autoinformation as the difference of conditional entropies highlights that it is maximized by an aggregated Markov chain with (i) a high number of approximately equiprobable states that maximize $H(y_{t})$, and (ii) a low uncertainty $H(y_{t+T} | y_{t})$ associated with the prediction of $y_{t+T}$ based on state $ y_{t}$.

\paragraph{Maximizing autoinformation as state-aggregation scheme.}

\begin{figure}[tb!]
  \centering
  \includegraphics[width=\linewidth]{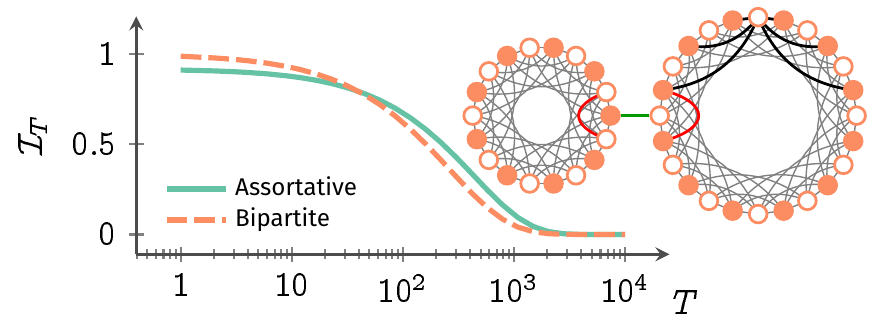}
  \caption{%
    Transition graph of a Markov chain with two alternative aggregations (inset schematic not of original size), an ``assortative'' split into two almost regular cyclic structures and a ``disassortative'', almost bipartite split.
    The black edges exemplify the linking pattern.
    Two additional edges (red) break the bipartite symmetry and one joins the two cycles (green).
    The autoinformation results are shown for $N=120+240$ nodes, with average degree $\langle k\rangle=10.02$.
    At short and long timescales, autoinformation is optimized in the bipartite or the assortative partition respectively.
  }\label{fig:sameB}
\end{figure}

The above discussion suggests maximizing the autoinformation $\mathcal{I}_T(h)$ over all possible state aggregations $h$ as a possible scheme to obtain a reduced order description.
Let us first explore the case in which we are given a desired cardinality $K$ of the aggregated state space $\mathcal Y$, i.e., we look for a partition of $\mathcal X$ into $K$ aggregation classes.
Denoting the space of all possible mappings to $K$ states as $\mathcal H_K$, we arrive at the following optimization problem to obtain a state aggregation $\hat{h}_T$:
\begin{equation}\label{eq:autoinformation_fixed_K}
  \hat{h}_T = \arg\! \max_{h \in \mathcal{H}_K} \mathcal{I}_T(h) = \arg\! \max_{h \in \mathcal{H}_K} I(y_{t+T}; y_t).
\end{equation}

To gain intuition, consider Eq.~\eqref{eq:autoinformation_fixed_K} when $x_t$ is a simple random walk process on an unweighted, undirected graph.
Then, finding an optimal state aggregation is equivalent to finding an optimal partition of the nodes.

Fig.~\ref{fig:sameB} displays a simple state transition graph of a Markov chain with two cyclelike subparts connected by a single link.
The cycles have even length and are constructed such that the graph is also almost bipartite.
Let us now consider the problem of finding a state aggregation of this chain in $K=2$ classes using autoinformation.
The autoinformation associated with both aggregation classes is qualitatively similar: at each time step the walker will likely both (\emph{i}) change node type with respect to the (almost) bipartite structure and (\emph{ii}) stay in the same cyclic structure.
At short timescales, $H(y_{t+T}|y_t) \approx 0$ for both structures and the $H(y_t)$ term in Eq.~(\ref{eq:autoinfo-bis}) dominates.
Accordingly, the bipartite partition, with slightly higher $H(y_t)\approx 1$, is preferred.
For longer timescales, however, the second term of Eq.~(\ref{eq:autoinfo-bis}) dominates and the two-cycle partition is preferred: there is a smaller probability of leaving each cycle than of changing the bipartite aggregation class (see Fig.~\ref{fig:sameB}).

\paragraph{Relationships to the degree corrected stochastic block model.}
A direct computation shows that optimizing Eq.~\eqref{eq:autoinformation_fixed_K} for $T=1$ is (coincidentally) equivalent to solving a maximum-likelihood estimation problem for the DCSBM with $K$ classes~\cite{Dasgupta2004,karrer2011dcsbm}.
More precisely, $\hat{h}_{T=1} = \arg\! \max \ell_\text{DCSBM}(\mathbf{A}) $, where $\mathbf{A} =[A_{ij}]\in \{0,1\}^{N\times N}$ is the binary adjacency matrix of the graph and $\ell_\text{DCSBM}$ is the log-likelihood function of the DCSBM with model parameters given by their maximum-likelihood estimates (for a formal proof see the SM).

The above result emphasizes that only paths of length 1 (edges) are essential to the likelihood function of the DCSBM, which derives from the mutual independence of edges in a DCSBM.
Interpreting the maximum-likelihood estimation for the DCSBM dynamically in terms of the autoinformation highlights this as a potential problem when fitting DCSBMs to graphs with long-range path structures.
Indeed, since optimizing autoinformation for $T=1$ for $K=2$ amounts to fitting a two-group DCSBM, Fig.~\ref{fig:sameB} shows that the two-cycle split would be missed when fitting such a graph via a DCSBM.
Our dynamical standpoint sheds light on the underlying issue: when only considering trajectories of length $1$, the description in terms of the bipartite structure will be preferred, because it offers a more balanced partition of the states into two equiprobable classes.
The specific path structure of this graph leads to slow mixing of the chain within and between the two cycles, and the assortative split is thus not apparent at $T=1$.
Stated differently, at timescale $T=1$ the bipartite switching is the dominant dynamical behavior of the Markov chain, and fitting a DCSBM to the state-transition graph \emph{correctly} captures this.

\paragraph{The importance of time-scales.}
Our approach also offers a way out of the above encountered dilemma: using $T\gg1$ shifts the focus to the slow modes of the dynamics, for which the assortative split into the two cyclic structures becomes clear.
The time parameter thus tailors the search to partitions that are dynamically relevant over longer timescales.
Unlike with many community detection methods featuring a resolution parameter, the time parameter does not offset a null model linearly, but acts \emph{nonlinearly} (see the SM, where we also prove the additional result that the optimal split into $K=2$ equiprobable aggregation classes of \emph{any} Markov chain tends to be either almost block diagonal (assortative) or almost bipartite (disassortative)).

\paragraph{How many aggregation classes?}
In many scenarios the number of aggregated states $K$ can be gleaned from prior knowledge and we thus have not discussed determining $K$.
In a scenario where $K$ is unknown, one would be tempted to  optimize the autoinformation over all partitions without a constraint on $K$ --- but this would yield the trivial state aggregation $y_t=x_t$ (see the SM).
This can be interpreted as data \emph{overfitting}: without constraints on $K$, the best aggregation corresponds to the original model, which trivially captures all available information.
To yield an aggregated description of size $K\le N$ when maximizing the autoinformation, we have to impose additional constraints on the state-aggregation mapping $h$.

For a given quality criterion such as the autoinformation two approaches are typically considered.
One would be to find state-aggregation mappings via Eq.~\eqref{eq:autoinformation_fixed_K} for a varying number of states $K \in \{1,\ldots, N\}$ and then select from among those solutions, e.g., using an elbow criterion (see the SM).
Here we follow another common approach by adding a complexity penalty to the objective function considered in Eq.~\eqref{eq:autoinformation_fixed_K}.
Optimizing the corresponding variational problem over all state-aggregation mappings, leads to an aggregated system that maximizes autoinformation while maintaining small complexity.
This general approach can be interpreted in terms of Occam's razor or a minimum description length (MDL) principle~\cite{grunwald2007minimum}.

For simplicity, we choose the description length necessary to describe the aggregated states of the aggregated state space as penalty term.
Specifically, we consider the regularized autoinformation with an entropy penalty:
\begin{equation}\label{eq:autoinfo_regularization}
  \mathcal{I}_{\beta,T}(h)
  = \mathcal{I}_T(h) - \beta H(h(x_t)),
\end{equation}
where $\beta$ is a Lagrange multiplier for the regularization term (see the SM for a discussion of these parameters).
However, our scheme is not bound to this specific complexity penalty and other regularization schemes such as the Aikake information criterion~\cite{akaike1974aic} or ideas from Bayesian statistics and MDL-based modeling~\cite{peixotoPRL2013parsimonious,grunwald2007minimum} may be considered.
The specific choice of entropy for capturing the complexity of the partition can be seen as a smooth generalization of $K$,  the number of classes, since $K$ equally-likely blocks translate into an entropy of $\log K$, while the entropy is also able to account for the size distribution of classes.

Like most combinatorial optimization problems, finding the aggregation that maximizes Eq.~\eqref{eq:autoinfo_regularization} is computationally difficult and we thus have to resort to a heuristic optimization.
Here we use an $\epsilon$-greedy optimization scheme akin to simulated annealing: starting from an initial partition, we stochastically loop over nodes and try to aggregate them with another class. If the regularized autoinformation improves we aggregate the node with the new class with probability 1; otherwise, we aggregate with probability $\epsilon\propto e^{- \Delta\mathcal I_{\beta T}/\tau}$, with $\tau$ a temperaturelike parameter that decreases along the maximization.
A detailed discussion is given in the SM, and a reference implementation is publicly \href{https://maurofaccin.github.io/aisa}{available}~\footnote{Code repository available at \url{https://maurofaccin.github.io/aisa}}.

\begin{figure}[!tb]
  \centering
  \includegraphics[width=\linewidth]{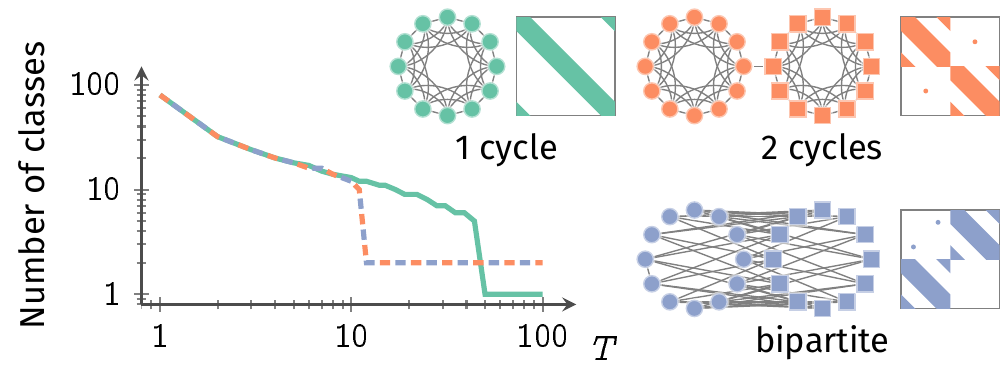}
  \caption{%
      Markov chains with natural timescales: a $k$-nearest neighbor  cycle with $N$ nodes (green), two cycles of $N/2$ nodes connected by a single edge (orange), and a bipartite graph with a single link breaking the symmetry (blue), see insets for schematics.
      The plots correspond to graphs with $360$ nodes with average degree $\langle k\rangle\approx 36$.
      The color of each line encodes the corresponding graph topology.
      At short timescales the maximization of the regularized autoinformation ($\beta=0.1$) tends to overfit the structure of these graphs with dense diagonal blocks (similar results hold for many community detection methods; see text).
      When increasing $T$, the algorithm finds the solution with the expected number of classes.
  }%
  \label{fig:plot_examples_betafix}
\end{figure}

\paragraph{Dynamical modules at short and long time-scales.}
The regularized autoinformation primarily provides a tool for state aggregations in Markov chains and dynamical data.
However, due to the connection with a maximum-likelihood estimation of a DCSBM, the regularized autoinformation also provides a dynamical view of certain model selection aspects under the DCSBM.

For concreteness, consider a random walk on a symmetric circular structure as the cycle of $N$ nodes connected to the $k$-nearest neighbors of Fig.~\ref{fig:plot_examples_betafix}.
For short timescales, it is sensible for dynamical model reduction to aggregate small patches of the cycle that are unlikely to be left by the walker after $T$ steps into aggregated states: the predictive power of such a fine-grained description outweighs the cost of the regularization term for most nonzero values of $\beta$.
In particular observe that maximizing Eq.~\eqref{eq:autoinfo_regularization} with $T=1$ leads to a nontrivial number of aggregated states as shown in Fig.~\ref{fig:plot_examples_betafix}.
By symmetry arguments, which patches of the cycle we use as aggregated states is irrelevant, and there is a large number of equivalent optimal aggregated system descriptions, corresponding to different (symmetric, regular) partitions of the cycle.

Interestingly, qualitatively similar results hold \emph{irrespective} of the regularization scheme used.
This explains why, e.g., inferring a DCSBM to such a cyclic graph with model selection via an MDL approach~\cite{peixotoPRL2013parsimonious} distinct from the regularization term used in Eq.~\eqref{eq:autoinfo_regularization}, results in a split into $22$ classes (for a more detailed discussion see the SM).
This ``overfitting'' behavior is in fact generic and can be observed with many other community detection algorithms, including modularity optimization~\cite{clausetPRE2014finding,Schaub2012fieldofview} and the map-equation framework~\cite{Rosvall29012008,Schaub2012}.
The issue is that while the graph structure can be compressed in terms of block structure with relatively small blocks, these blocks are less relevant for the long-term dynamics.

As seen in Fig.~\ref{fig:plot_examples_betafix}, for Markov chains with sparse state-transition graphs with long-range path structures, this mismatch between clusters defined via one-step block connectivity ($T=1$) and clusters capturing the long-term behavior can be quite pronounced.
Indeed, the regularized autoinformation for short times is typically optimized by choosing a relatively large number of aggregated states, while the dynamically planted class structure is only found for larger $T$.
This short time behavior of the regularized autoinformation is again mirrored by the MDL-based inference of DCSBMs or the map equation, which both fail to find the dynamically meaningful partition for long timescales for all the graphs shown in Fig.~\ref{fig:plot_examples_betafix}: the inference of the DCSBM using the MDL approach in~\cite{peixotoPRL2013parsimonious} yields around 22 classes in all cases; the map equation provides 7, 10 or 4 aggregation classes for the three scenarios, respectively.
While in this case spectral methods, such as GenPCCA~\cite{fackeldey2018spectral} can resolve the relevant structure, they may fail when intermediate or short timescales are of interest.
More in-depth comparisons can be found in the SM, where we also describe a synthetic model class (a graph ensemble) that displays the behavior observed here.
We emphasize that changing the parameters $\beta$ or $T$ is in general \emph{not equivalent} (see the SM).

\begin{figure}[!htb]
  \centering
  \includegraphics[width=\linewidth]{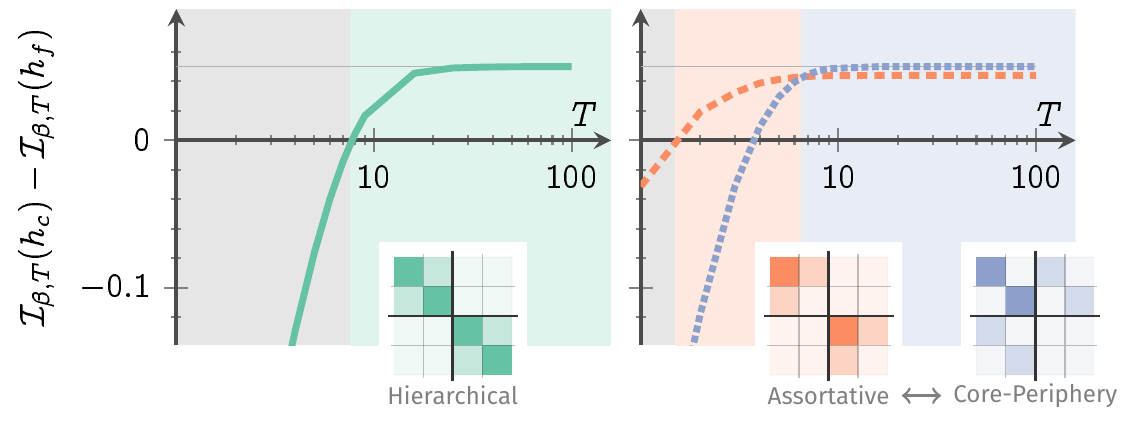}
  \caption{State aggregations for Markov chains with hierarchical timescales on an SBM.
    Left: We plot the difference in the regularized autoinformation between a fine state-aggregation $h_f$ into the four planted aggregation classes, and a coarse two-class state aggregation $h_c$, for a hierarchical state-transition graph of a Markov chain (inset).
    The plot shows that at longer timescales (green shade) $\mathcal{I}_{\beta,T}(h_c)> \mathcal{I}_{\beta,T}(h_f)$ and hence the coarser aggregation is preferred over the fine aggregation, which is preferred at shorter timescales (gray shade).
    Right: Difference in the regularized autoinformation between the two-class split $h_c$, describing either a core-periphery (orange) or an assortative (violet) aggregation and the underlying planted aggregation into four classes ($h_f$).
    The four-class partition has a higher autoinformation than the two-class split at short timescales (gray shade).
    The assortative partition has highest autoinformation for middle range timescales (orange shade) and the core-periphery partition is preferred at longer timescales (blue shade).
    All graphs consist of $400$ nodes and expected average degree $\langle k\rangle=15$, with $\beta=0.05$.
  }%
  \label{fig:plotexamples_hier}
\end{figure}

\paragraph{Hierarchical aggregation of Markov chains with multiple timescales.}
In a hierarchical Stochastic Block Model (see Fig.~\ref{fig:plotexamples_hier}, left), for any given value $\beta >0$, the finer structure is typically \emph{preferred} at lower values of $T$ where the walker dynamics are confined to the local class.
Higher values of $T$ allow the walker to visit larger portions of the network, and coarser partitions gain importance.

Consider now two alternative hierarchical aggregations (core periphery vs assortative) of an initial aggregation into four classes (see Fig.~\ref{fig:plotexamples_hier}, right).
While the same four-class structure is preferred at short timescales, the two-class assortative and the core-periphery structures are preferred at medium and longer timescales.
Although for two equally sized classes (in terms of entropy), the optimal aggregation is either assortative or disassortative, here, the core and periphery are of different sizes in terms of the probability of the presence of the walker.
In particular, the regularization term $\beta H(y_t)$ in Eq.~(\ref{eq:autoinfo_regularization}) favors the core-periphery split.
This effect dominates at large timescales, where the autoinformation converges toward zero.
This is intrinsic to what our regularization term in Eq.~(\ref{eq:autoinfo_regularization}) considers to be a `small' or `simple' model, and other choices of regularization may lead to different results.
Similar trade-offs were observed for SBMs and DCSBMs in~\cite{karrer2011dcsbm, peel2017ground_truth} when considering core-periphery or assortative structures.

\begin{figure}[!tb]
  \centering
  \includegraphics[width=\linewidth]{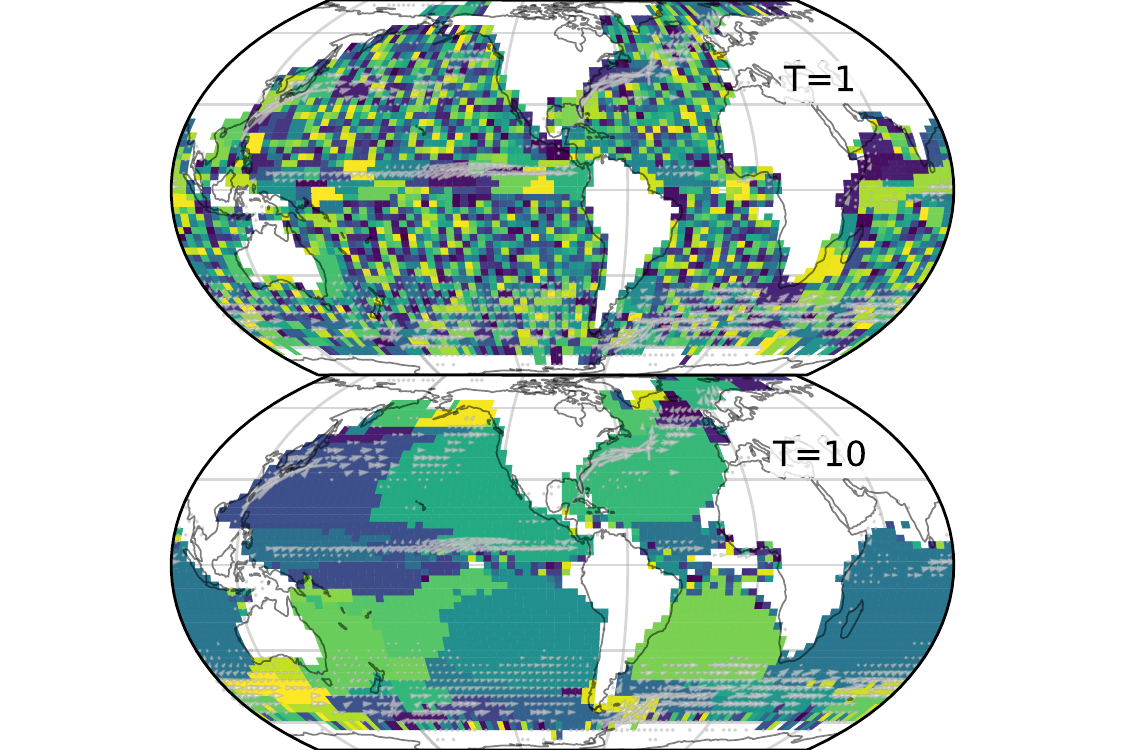}
  \caption{State aggregation of ocean currents.
    The above maps compare two partitions induced by aggregating the states of the ocean currents according to the regularized autoinformation for short timescales (top) or longer timescales (bottom) with $\beta=0.5$.
    At shorter timescales a higher number of classes is found.
    At longer timescales, the aggregation classes reveal well-known features of global ocean dynamics such as the Antarctic Circumpolar Current, subtropical gyres and, in general, a marked separation of the polar, midlatitude and equatorial regions.
    The quiver plot overlay displays the average drifter's velocity.
    Each time step $t$ corresponds to 16 days.
  }%
  \label{fig:oceans}
\end{figure}

\paragraph{The system of ocean surface currents.}
Let us now showcase how one can use the autoinformation as a tool to analyze dynamical data.
The Global Drifters Program\footnote{\href{http://www.aoml.noaa.gov/phod/gdp/index.php}{Global Drifter Program: http://www.aoml.noaa.gov/phod/gdp/index.php}} tracks drifter buoys on the surface of all oceans.
The dynamics of the drifters is a proxy for the global system of surface currents, i.e., water masses moving between different areas of the ocean surfaces.

Using the regularized autoinformation, we identify macro areas that optimally aggregate the drifter dynamics.
We find that the temporal dimension of the kinetics strongly influences the outcome.
For short timescales, the aggregation classes correspond to small geographic patches of ocean surface that become larger where currents are stronger and steadier, e.g., along the equator (see Fig.~\ref{fig:oceans} top).
For timescales closer to the expected time for a drifter to cross an ocean, larger geographic patches are found.
These encompass all major ocean gyres (see Fig.~\ref{fig:oceans} bottom) separating equatorial, subtropical and boreal regions.
The northern and southern Pacific are subdivided into western and eastern parts that belong to the same large-scale circulation pattern, but represent different areas of surface convergence and are located around so-called garbage patches~\cite{young2009PLOSONE, howell2012pollution, Lebreton2018}.

Recently, the ocean currents have been clustered in dynamical domains by analyzing a long-term simulation of the barotropic vorticity equation, and applying a simple $k$-means algorithm on the magnitude of the different terms contributing to the vorticity dynamics~\cite{sonnewald2019ocean_Kmeans}.
This dynamics is only partly comparable to the drifter dynamics as it involves not just surface currents but an average over all ocean depths.
It is nonetheless interesting to compare the outcomes, which share many features (see the SM for these comparisons).
However, a key difference is that while the $k$-means method~\cite{sonnewald2019ocean_Kmeans} can lead to geographically disconnected patches, scattered across the globe, our method finds spatially connected classes and is moreover completely data-driven, using only the multiscale dynamical analysis of empirical trajectories.

\begin{acknowledgments}
    \small
M.T.S. received funding from the European Union’s Horizon 2020 research and innovation program under the Marie Sklodowska-Curie grant agreement no. 702410 and the Ministry of Culture and Science (MKW) of the German State of North Rhine-Westphalia (``NRW Rückkehrprogramm'').
M.F. was partially funded by Innoviris grant no. D1.31402.007-F.
J.C.D. was partially funded by the Flagship European Research Area Network (FLAG-ERA) Joint Transnational Call “FuturICT 2.0”.
We warmly thank Leto Peel and Eric Deleersnijder for fruitful discussions.
\end{acknowledgments}

\bibliography{biblio_arxiv}

\appendix
\setcounter{figure}{0}
\renewcommand\thefigure{SM\arabic{figure}}
\setcounter{equation}{0}
\renewcommand\theequation{SM\arabic{equation}}

\begin{center}
  \Large\textbf{Supplemental Material}
\end{center}
\section{Autoinformation and its properties}
In this section we describe a number of properties of the autoinformation in more detail.

\subsection{Relationship to DCSBM for one-step random walk dynamics}%

Here we show that the log-likelihood for a given partition of an symmetric, binary network is (up to factors that are non-essential for its optimization) equivalent to the one-step autoinformation $\mathcal I_1(h)$ of the corresponding aggregated dynamics of a simple random walk on the network.

We recall that the maximization of the log-likelihood for the DCSBM~\cite{newman2015generalized} for a partition into $K$ blocks corresponds to the minimization of:
\begin{equation*}
  \mathcal S = E - \sum_{ij} e_{ij} \log \frac{e_{ij}}{e_i e_j},
\end{equation*}
where $e_{ij}$ is the sum of the adjacency matrix entries connecting nodes in block $i$ to nodes in block $j$, $e_i = \sum_j e_{ij}$ is the sum of links attached to nodes in class $i$, and $E$ is a constant (see, e.g., ~\cite{peixotoPRL2013parsimonious}).
By expanding the logarithm, this can be rewritten as:
\begin{equation*}
  \mathcal S =
  E - \sum_{ij} e_{ij} \log e_{ij} + 2 \sum_i e_i \log e_i.
\end{equation*}

Now, since $\sum_{ij} e_{ij} = \sum_i e_i = 2m$ is twice the number of edges in the network we can rewrite the above quantities as:
\begin{align*}
  \sum_{ij} e_{ij} \log e_{ij} = &
  2m \sum_{ij} \frac{e_{ij}}{2m} \log \frac{e_{ij}}{2m} + 2m \log(2m), \\
  \sum_i e_i \log e_i =          &
  2m \sum_i \frac{e_i}{2m} \log \frac{e_i}{2m} + 2m \log(2m),
\end{align*}
Plugging these equations into the above expression gives:
\begin{align*}
    \mathcal S =&\;
    E - 2m \sum_{ij} \frac{e_{ij}}{2m} \log \frac{e_{ij}}{2m} \\  &+ 4m \sum_i \frac{e_i}{2m} \log \frac{e_i}{2m}+ 2m \log(2m).
\end{align*}

Finally observe that for a stationary random walk on a symmetric, binary network we will have $H(y_t) = H(y_{t+1})$ (by stationarity).
Further the occupation probabilities of the blocks are $p(y_t = i) = \frac{e_{i}}{2m}$, and transition probabilities between the blocks are given by $p(y_t = i, y_{t+1} = j) = \frac{e_{ij}}{2m}$.
We can thus assert by direct computation that:
\begin{align*}
  \mathcal S &=
  E + 2m \left[H(y_t, y_{t+1}) - 2 H(y_t) +\log(2m) \right]\\
  &=
  E - 2m \left[H(y_t) - H(y_{t+1}| y_t) -\log(2m) \right]\\
  &=
  E - 2m \left[\mathcal I_1(h) - \log(2m) \right].
\end{align*}
Hence, minimizing $\mathcal S$, i.e., maximising the likelihood of the DCSBM with parameters given by their maximium likelihood estimates, corresponds to maximizing the autoinformation for $T=1$ if the network is symmetric and binary.

A number of points of the above result are worth emphasizing.
While the DCSBM is a generative network model, maximising the autoinformation does not impose a generative process of the data and can be applied to any dynamical process with a discrete state space.
For instance, the autoinformation can be computed without modification for a Markov process defined by a random walk on a \emph{weighted} network, or a set of trajectory data without an explicitly defined network.
In contrast, the DCSBM is a priori specified only for unweighted networks.
This corresponds to the fact that edges are all independent in the DCSBM and, hence, only paths of length one (i.e., edges) are essential to its likelihood function.

Note that autoinformation beyond $T=1$ or beyond simple random walks on unweighted graphs is not, to our best knowledge, the likelihood function of a generative model, thus cannot be maximised in general by a statistical inference technique.

\subsubsection{Over-fitting of non-block structures using a DCSBM.}

The aim of algorithms based on stochastic block models or their variants is to decompose the adjacency matrix into groups which are `simple' (e.g.\ with density that is approximately constant or proportional to a degree distribution).
This offers a `dictionary' of patterns that is universal, in that it can eventually fit any network.
Nevertheless, such patterns, applied to cycle-like graphs such as depicted on Fig.~\ref{fig:plot_examples_betafix}, will generate a large number of blocks to fit the \emph{banded shape} of the adjacency matrix.
Thus the dictionary of the DCSBM is not adapted for an efficient description of the cyclic structures, which are `simple' in another fashion.
The situation is similar in some regard to the approximation theorems in numerical analysis: we know that any continuous real-valued function on the interval can be approximated arbitrarily well by polynomials (Weierstrass theorem) or by sines and cosines (Fourier decomposition), or by many other basis functions, but some functions are more efficiently approximated by polynomials and some others by a truncated Fourier decomposition.
Here, with similar arguments, one can assert that in some cases (i.e., those considered in Fig.~\ref{fig:plot_examples_betafix}) a block model is an inefficient basis to describe structures such as a banded adjacency matrix, while the `dictionary' offered by higher time scales $T>1$ is more appropriate.

\subsection{Extreme values of (regularized) autoinformation and data processing inequalities}

In this section we consider the state aggregation mappings with maximal (regularized) autoinformation.

Observe that for any two random variables $X$ and $Y$ and deterministic maps $f$
 and $g$, it is well known that $I(f(X);g(Y)) \leq I(X;Y)$.
This is a form of the data-processing inequality~\cite{cover2012elements}.
We apply this data-processing inequality to the autoinformation of a Markov chain ($I(x_{t+T};x_t)$), and its aggregation, $I(y_{t+T};y_t)$, through the aggregation map $y_t=h(x_t)$, to obtain:
\begin{equation*}
I(y_{t+T};y_t) \leq I(x_{t+T};x_t).
\end{equation*}
Thus the aggregation maximizing the autoinformation is the trivial aggregation $y_t=x_t$, with $h$ being the identity map.

For the regularized autoinformation
\begin{equation*}
\mathcal{I}_{\beta,T}(h)=I(y_t;y_{t+T}) - \beta H(y_t)
\end{equation*}
with $\beta=1$, we see that it reduces to $-H(y_{t+T}|y_t)$, which takes its maximal value  of zero for the trivial constant aggregation $h$ (all states of $\mathcal X$ being mapped to the single element set $\mathcal{Y}=\{y\}$).

\subsection{Analysis of the (regularized) autoinformation for limiting cases}

In the following we analyze how the autoinformation maximization behaves when considering two aggregated states, short or long time-scales.
To remove the effect of regularization, when comparing two different partitions, we compare aggregations of same complexity $H(y_t)$.

\subsubsection*{Two aggregated states}

We consider the ideal case of $K=2$ aggregated states denoted by $y=1$ and $y=2$ of same occupation probability $p(y_t=1)=p(y_t=2)=1/2$ (equivalently, $H(y_t) = 1$).
The joint probabilities on successive states fulfill the following set of equalities:
\begin{align}
p(y_t=y_{t+T}=1)+p(y_t=1,y_{t+T}=2)&=1/2, \label{eq:2state-1} \\
p(y_t=y_{t+T}=1)+p(y_t=2,y_{t+T}=1)&=1/2,  \label{eq:2state-2}  \\
p(y_t=2,y_{t+T}=1)+p(y_t=y_{t+T}=2)&=1/2,  \label{eq:2state-3} \\
p(y_t=1,y_{t+T}=2)+p(y_t=y_{t+T}=2)&=1/2.  \label{eq:2state-4}
\end{align}

We define the quantity $p_\text{leak,$T$}=p(y_t \neq y_{t+T})$, which, in the simple case of two classes, is $p(y_t=1,y_{t+T}=2)+p(y_t=2,y_{t+T}=1)$.
The above equalities, subtracting Eq.~(\ref{eq:2state-4}) from Eq.~(\ref{eq:2state-3}) and Eq.~(\ref{eq:2state-2}) from Eq.~(\ref{eq:2state-1}), enable us to write:
\begin{align*}
p(y_t=1,y_{t+T}=2)&=p(y_t=2,y_{t+T}=1)=\frac{p_\text{leak,$T$}}{2}, \\
p(y_t=1,y_{t+T}=1)&=p(y_t=2,y_{t+T}=2)=\frac{1-p_\text{leak,$T$}}{2}.
\end{align*}

 Note that this calculation implies, in particular, that in this case the aggregated Markov chain is reversible even when the original Markov process on $\mathcal{X}$ is not.

From the above calculations we conclude that
\begin{equation*}
H(y_{t+T}|y_t)=H(\mathds{1}_{y_t \neq y_{t+T}})=\mathcal{H}(p_\text{leak,$T$}),
\end{equation*}
where $\mathcal{H}(x)$ denotes the Shannon entropy function of a probability $x$, $\mathcal{H}(x)=-x \log x - (1-x) \log (1-x)$; and the notation $\mathds{1}_{S}$ stands for the indicator variable of the event $S$, taking value $1$ if the event $S$ is realized and $0$ otherwise.

Therefore the autoinformation for an aggregation into two classes with same complexity can be written as:
\begin{equation*}
I(y_{t+T};y_t)= H(y_{t+T}) - H(y_{t+T}|y_t)=1-\mathcal{H}(p_\text{leak,$T$}).
\end{equation*}

In other words, the autoinformation is in this case determined by the leak probability $p_\text{leak,$T$}$.
It is maximized when $p_\text{leak,$T$}$ is either as low as possible or as large as possible.
The former case can be identified as an `assortative' partition of the original Markov chain (relatively to time scale $T$) and the latter, as a `disassortative' partition of the original Markov chain (relatively to time scale $T$).

This terminology generalizes the usual notion of assortativity coefficient in the following way.
Considering the random walk on an binary symmetric network with two classes of nodes, Newman's binary assortativity coefficient~\cite{newman2002assortative} is also in one-to-one relationship with $p_\text{leak,$1$}$, for $T=1$ step, and takes high values (close to $+1$) for $p_\text{leak,$1$}$ close to $1$, and low values (close to $-1$) for $p_\text{leak,$1$}$ close to $0$.

\subsubsection*{Short time scales}
Let us characterize the autoinformation~\eqref{eq:autoinfo-bis} of an arbitrary  aggregated Markov chain for short time-scales.
To make our analysis more meaningful, we will here switch our focus to a continuous-time Markov chain, as it will enable us to consider the limit of the step size (respectively the transition rate) going to zero.

A continuous-time Markov chain on state space $\mathcal{X}$ is in a state $i$ at the real time instant $t$, and makes a transition in the infinitesimal interval $[t,t+dt]$ to another state $j$ with probability $L_{ij}dt$, for some rate $L_{ij} \geq 0$.
Introducing the quantity $L_{ii}=-\sum_j L_{ij} \leq 0$, the Markov chain remains in state $i$ throughout $[t,t+dt]$  with probability $1+L_{ii}dt$.
Arranging the coefficients $L_{ij}$ into a Laplacian-like matrix $L$ (with rows summing to zero), we can describe the evolution of state probabilities of each state with the following master equation:
\begin{equation*}
  \dot{p}(t)=p(t) L
\end{equation*}
where $p_i(t)$ is the probability of the Markov chain to be in state $i$ at time $t$, and $p(t) = [p_1(t),\ldots p_N(t)]$ is the row-vector collecting all probabilities $p_i(t)$.
The above dynamics lead to a state-transition equation of the form:
\begin{equation*}
    p(t+T) = p(t)\text{exp}(LT),
\end{equation*}
where $\text{exp}(\cdot)$ is the matrix exponential function.

We now consider the aggregated process on $\mathcal{Y}$. Let us consider the indicator variable $\mathds{1}_{y_t \neq y_{t+T}}$ taking value 1 if a change of aggregated state occurs or 0 otherwise. This is again the \emph{leak probability} or \emph{escape probability} $p_\text{leak,$T$}$, as discussed in the previous section.
For short time-scales $T\rightarrow 0$, we have $\exp(LT) \approx I + LT$, and therefore
$$p_\text{leak,$T$} \approx \sum_{i,j \in \mathcal{X}:h(i)\neq h(j)} p(x_t=i) L_{ij} T=\mathcal{O}(T).$$

For the derivations below it is useful to remember that the Shannon binary entropy function $\mathcal{H}(x)=-x \log x - (1-x) \log (1-x)$ scales as $-x \log x$ for $x \rightarrow 0$.
For instance $H(\mathds{1}_{y_t \neq y_{t+T}})=\mathcal{H}(p_\text{leak,$T$})$ scales as $p_\text{leak,$T$} |\log T|$ for small $T$, since $\log \sum_{i,j \in \mathcal{X}:h(i)\neq h(j)} p(x_t=i) L_{ij}$ is a constant, while $|\log T| \to +\infty$.

 We can now write
\begin{align}
  H(y_{t+T}|y_t) & =H(y_{t+T},\mathds{1}_{y_t \neq y_{t+T}}|y_t) \nonumber      \\
                 & =H(\mathds{1}_{y_t \neq y_{t+T}}|y_t)+H(y_{t+T}|y_t,\mathds{1}_{y_t \neq y_{t+T}}) \label{eq:si1jump} \\
                 &\approx H(\mathds{1}_{y_t \neq y_{t+T}}|y_t) \label{eq:si2jump}  \\
                 & =\sum_{k \in \mathcal{Y}} p(y_t=k) H(\mathds{1}_{y_t=k \neq y_{t+T}})  \nonumber \\
                 	&=\sum_{k \in \mathcal{Y}} p(y_t=k) p(y_t=k \neq y_{t+T}) |\log T| \nonumber \\
                & = p_\text{leak,$T$} |\log T|. \nonumber
\end{align}
We derive Eq.~(\ref{eq:si1jump}) from the chain rule for joint entropy $H(X,Y|Z)=H(X|Z)+H(Y|X,Z)$ (for arbitrary random variables $X,Y,Z$).  We derive Eq.~(\ref{eq:si2jump}) by observing that the first term in Eq.~(\ref{eq:si1jump}) turns out to scale as $T|\log T|$, whereas the second term scales as $p_\text{leak,$T$}$, thus as $T$, which is dominated by $T|\log T|$ for $T \rightarrow 0$.

In conclusion, in the short time limit, the dominant term of the conditional entropy is the \emph{leak probability} from the aggregated states, up to a factor  $\log T$.
Accordingly, the autoinformation $I(y_t;y_{t+T})= H(y_t)-H(y_{t+T}|y_t)$ in a continuous-time Markov process with a fixed $H(y_t)$ is maximized for $T\rightarrow 0$  by choosing a state aggregation such that the flow of the process gets trapped within each block.

In the case of a continuous-time random walk on an symmetric graph, $L$ is the usual Laplacian, and  the leak probability from the aggregated states is essentially given by the `cut size' between the blocks (`communities') of  nodes in the graph, i.e.\ the total weight of edges standing between the aggregation classes.
A number of `node partitioning' or `community detection' methods aim at minimizing this cut size, regularized with a constraint or entropy-like cost promoting a nontrivial number of equal-size blocks.
This strategy underlies most edge-counting methods such as conductance, normalized cuts, ratio cuts, modularity, Potts model and linearized Markov stability.
See~\cite{delvenne2013stability} for references, discussion and detailed arguments.
Maximizing the short-term autoinformation is essentially identical to what all these methods implement, up to the choice of regularization strategy, and will accordingly yield an `assortative' aggregation, minimizing the leak from the aggregated states.
These type of `assortative' partitions are also the foundation for most time-scale separation techniques on Markov chains.
As a concrete example, assume the Laplacian  $L$ is written as $L_0+\epsilon L_1$, where $L_0$ is a block-diagonal Laplacian matrix  describing the union of decoupled Markov chains, and the rows of $L_1$ sum to zero.
Then it it is a classic result~\cite{simon1961aggregation} that for $\epsilon\rightarrow 0$, the Markov chain aggregated along the diagonal blocks of $L_0$ is indeed approximately Markovian, and can therefore be safely replaced by a Markovian approximation.
Note that this trade-off between the Markov property and the dynamical predictability is precisely what is captured by the autoinformation, see Eq.~(\ref{eq:mutual_info}).

\subsubsection*{Long time scales}

Let us now consider the autoinformation for long time-scales, for which we revert to a discrete-time formulation.

Note that we can rewrite the autoinformation from the form in Eq.~\eqref{eq:autoinfo-bis} as:
\begin{equation}
  I(y_{t+T};y_t)=\left\langle \log \frac{p(y_{t+T},y_t)}{p(y_t)p(y_{t+T})} \right\rangle,
\end{equation}
where $\langle\cdot\rangle$ denotes the expectation, taken over the joint the distribution $p(y_{t+T},y_t)$.
Since for a symmetric graph the chain will converge towards a stationary state, the autoinformation will be close to zero for large $T$ for a mixing Markov chain.

We can therefore approximate the above expression using a Taylor expansion for the natural logarithm as:
\begin{align*}
I(y_t;y_{t+T}) &\approx \left\langle  \frac{p(y_t,y_{t+T})}{p(y_t)p(y_{t+T})}   -1\right\rangle\\
 &= \left\langle  \frac{p(y_t,y_{t+T})}{p(y_t)p(y_{t+T})}  \right\rangle -1\\
	&=\sum_{y_t,y_{t+T}} \frac{p^2(y_t,y_{t+T})}{p(y_t)p(y_{t+T})}-1.
\end{align*}

This can be further developed by estimating $p^2(y_t,y_{t+T})$ for $T \rightarrow \infty$. The Markov chain on the state space $\mathcal{X}$ (which we assume to be ergodic and mixing) is described by the one-step transition matrix $P$, which appears in the master equation:
$$
p(t+1)=p(t)P
$$
The transition matrix $P$ can be decomposed spectrally as  $P=\bm{1}p + \lambda v u + \ldots$, where $p$ is the stationary row-vector of occupation probabilities on $\mathcal{X}$, $\lambda$ is the second eigenvalue in magnitude, $u$ is the corresponding (column) right eigenvector ($Pv=\lambda v$) and $u$  is the (row) left eigenvector.
These eigenvectors satisfy $u \bm 1 = 0  = p v$, and $uv=1$.

Let us now suppose for convenience that $\lambda$ is real and unique.
For large $T$, we find that $P^T\approx\bm{1}p + \lambda^T v u$ as the powers of all other eigenvalues decay faster than $\lambda^T$.
Accordingly we can write
$$p(x_t,x_{t+T}) \approx p(x_t)p(x_{t+T}) + \lambda^T v(x_t) u(x_{t+T}).$$
Passing to the aggregated states, we obtain:
$$p(y_t,y_{t+T}) \approx p(y_t)p(y_{t+T}) + p(y_t) \lambda^T pv(y_t) u(y_{t+T}).$$
Here $pv(y_t)$ denotes the sum of all entries  $p(x_t)v(x_t)$ for all states $x_t$ aggregated to $y_t$, and $u(y_{t+T})$ is the sum of $u(x_{t+T})$ over all $x_{t+T}$ aggregated to $y_{t+T}$.

Thus we obtain the following approximation:
\begin{align*}
I(y_t;y_{t+T}) &\approx \sum_{y_t,y_{t+T}} \frac{p^2(y_t,y_{t+T})}{p(y_t)p(y_{t+T})} -1 \\
&\approx \sum_{y_t,y_{t+T}} \lambda^{2T} \frac{pv(y_t)^2 u(y_{t+T})^2}{p(y_t)p(y_{t+T})}\\
&= \lambda^{2T}   \sum_{y_t} \frac{pv(y_t)^2}{p(y_t)}    \sum_{y_{t+T}} \frac{  u(y_{t+T})^2}{p(y_{t+T})}
\end{align*}

In conclusion, the aggregation with highest autoinformation (in combination with a regularization criteria) in the large $T$ limit can be determined from the second left and right eigenvectors of $P$.
Thus the optimal aggregation can be determined exactly by a spectral algorithm (i.e.\ an algorithm exploiting the dominant eigenvectors $u$, $v$).
A complete description of such a spectral algorithm, whose details depend on the chosen regularization, is beyond the scope of this article.
To build the intuition for such an algorithm, we observe that in the case of a reversible Markov chain (for instance a simple random walk on a symmetric network), $p(x_t)v(x_t)=u(x_t)$ and the usual idea of splitting between positive and negative values of $u$ may  typically offers a good aggregation into two states, with high sum-of-squares $\sum_{y_t} \frac{u{(y_t)}^2}{p(y_t)}$.
This split will be an `assortative' splitting of $\mathcal{X}$  (if $\lambda >0$), or `disassortative', almost-bipartite splitting (if $\lambda < 0$).
This agrees with the general result for two-state aggregation above.

\subsection{Differences between time-scale \texorpdfstring{$T$}{T} and regularization parameter \texorpdfstring{$\beta$}{beta}}
Here we expand on the different roles of the time-scale parameter $T$ and the regularization parameter $\beta$.

The definition of the (unregularized) autoinformation in Eq.~\eqref{eq:autoinfo} contains only a time-scale $T$.
While this parameter changes the transition properties of the Markov process, unless a regularization term is introduced, optimizing the autoinformation in the space of all partitions will always lead to the trivial partition in which all nodes are in their own group.
We add this regularization term, which is here chosen to be the entropy of the aggregated state space, with a scalar multiplier $\beta$.
The parameter $\beta$ thus regulates the influence of the regularization akin to a Lagrange multiplier.

Changing $T$ or $\beta$ has a markedly different effect on the optimization landscape of the regularized autoinformation Eq.~\ref{eq:autoinfo_regularization}.
The parameter $\beta$ provides a \emph{linear} scaling of the entropic cost term which is independent of the detailed graph structure but merely depends on the size of the aggregated states in terms of the aggregated degrees.
In contrast, the temporal parameter $T$ acts in a \emph{non-linear} way and directly changes the time-scale of the dynamics.
In particular, the effect of $T$ depends on the details of the path structure of the underlying graph and cannot be understood by some local statistics such as the degree sequence.
For a related discussion in the context of dynamics-based graph embeddings, see also~\cite{Schaub2019}.

\begin{figure}[htb!]
  \centering
  \includegraphics[width=0.5\linewidth]{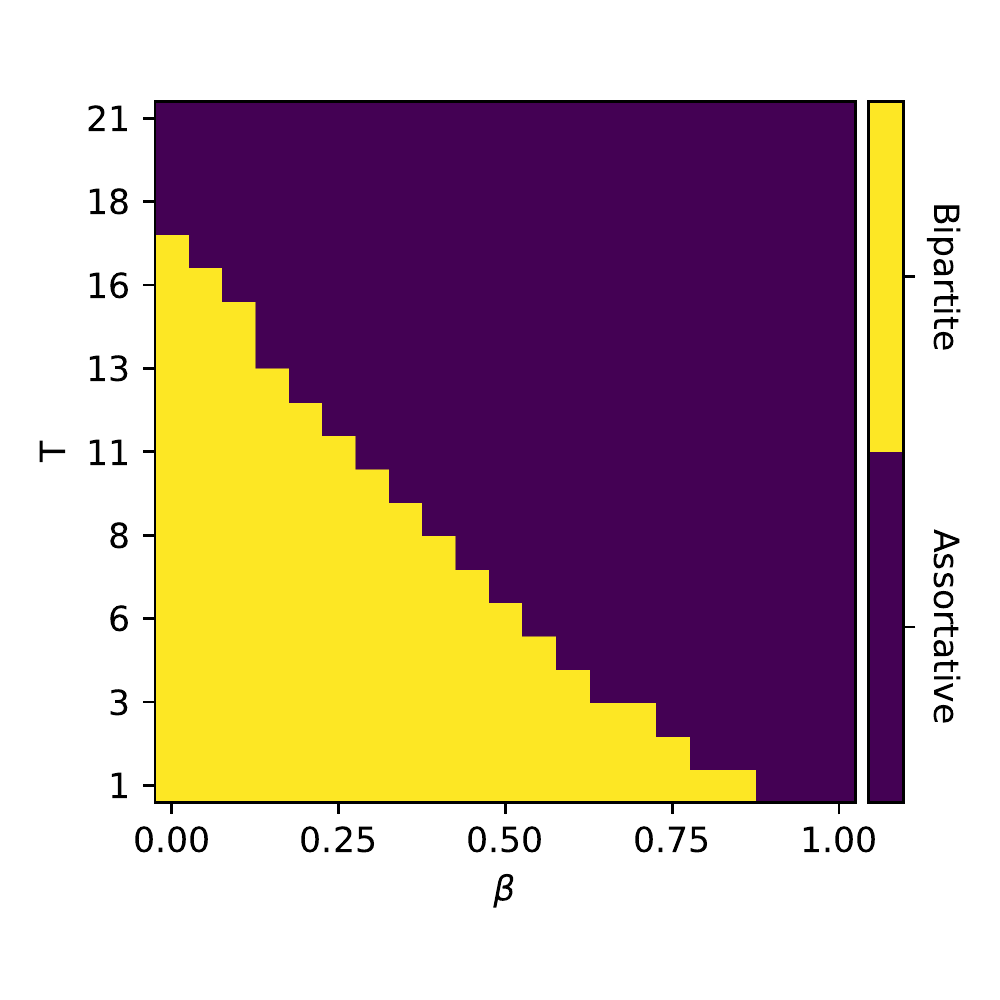}
  \caption{%
    The effect of $\beta$ and $T$ for optimal state-aggregation for a network with two alternative aggregations, as plotted in Fig.~\ref{fig:sameB} of the main text.
  The plot shows the parameter regime in which the preferred state aggregation is either the ``assortative'' split into two cycles or the ``disassortative'' almost bipartite split (see Fig.~\ref{fig:sameB}).}%
  \label{fig:parameter_space}
\end{figure}

Nonetheless, the parameters $\beta$ and $T$ may in some cases have a similar effect on the granularity of partitions obtained from optimizing Eq.~\eqref{eq:autoinfo_regularization}.
For instance, in Fig.~\ref{fig:parameter_space} we revisit the example network of Fig.\ref{fig:sameB}, in which both an almost bipartite split (for short time-scales) and a split into two cyclic structures (for long time-scales) provide a good aggregated description of the dynamics.
Accordingly, at low values of $T$, the bipartite partition is selected when optimizing the regularized autoinformation, but for large values of $T$, the split into two rings is obtained.
The same effect can here be obtained by fixing $T$ and regulating $\beta$: for a small $\beta$ the almost bipartite split is preferred and the split into two cycles is preferred from large $\beta$.

\begin{figure}[htb!]
  \centering
  \includegraphics[width=\linewidth]{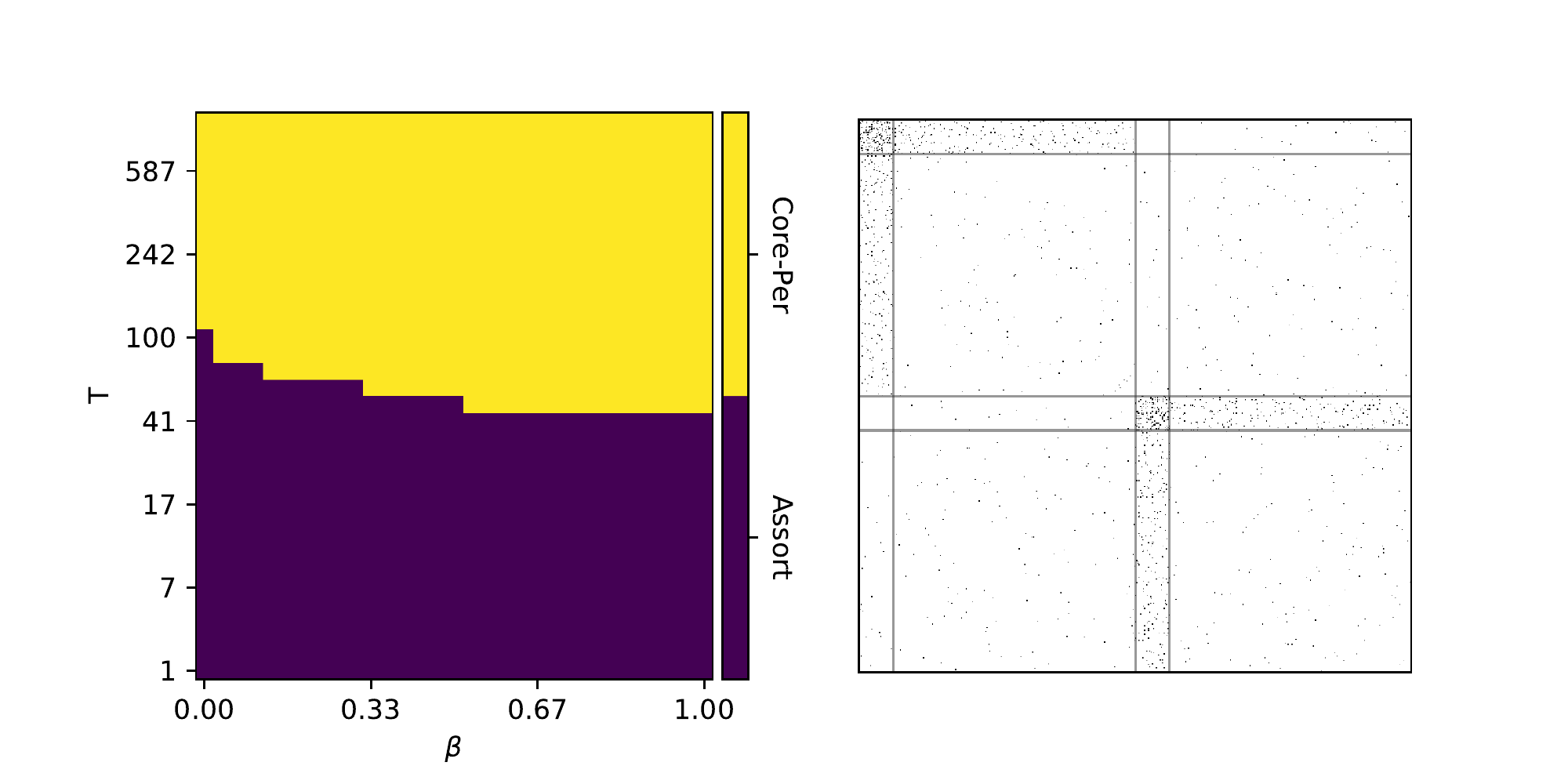}
  \caption{%
      The effect of $\beta$ and $T$ for optimal state-aggregation for a network with two alternative aggregations: a core-periphery and an assortative block-structure (see also Fig.~\ref{fig:plotexamples_hier}).
      The plot shows the parameter regime in which each of these partitions corresponds to the optimal state aggregation.
}\label{fig:parameter_space_cpass}
\end{figure}

However, as our next example illustrates, the effect of $\beta$ and $T$ on the chosen partition is indeed different, in general.
In Fig.~\ref{fig:parameter_space_cpass}, we consider a random walk on a network that may be partitioned in terms of a core-periphery structure, as well as a block-diagonal (``assortative'') partition.
As Fig.~\ref{fig:parameter_space_cpass} illustrates, the effect of $T$ and $\beta$ is clearly different in this case.
For small $T$ there is no setting of $\beta$ under which the core-periphery structure would lead to a more accurate description according to the regularized autoinformation.
For large $T$, however, the core-periphery split is always preferred.
This illustrates that the time-scale parameter $T$ may in some cases be necessary to find certain dynamically relevant structures.

\section{Details on the ocean drifters experiment}

\begin{figure}[th]
  \centering
  \includegraphics[width=\linewidth]{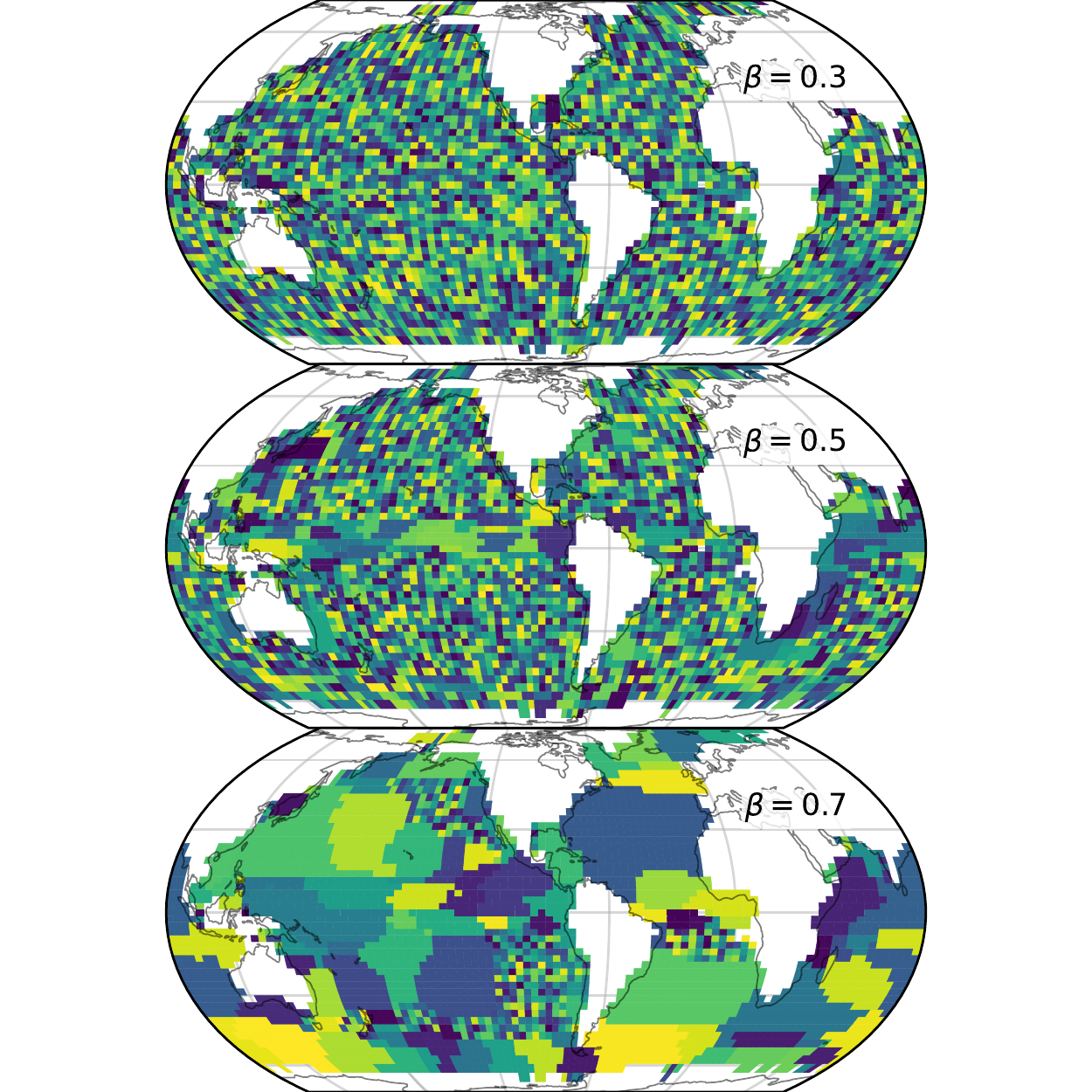}
  \caption{%
    Partitioning the oceans according to the found aggregation classes of the ocean drifters for short time-scales.
    For any value of the regularization parameter $\beta$ and $T=1$ the resulting partition comprises only small/local aggregation classes.
  }\label{fig:drifters-short}
\end{figure}

We divide the Earth surface in a grid of equal-area cells.
The Equator is divided into 100 equally spaced intervals of $3.6^\circ$ and the meridians into 50 intervals with varying length such that all cells have the same area under the assumption that the Earth is a perfect sphere.
Each cell represents a node of the graph and an edge is added between cell $i$ and cell $j$ if a drifter visited the cell $i$ at any time $t$ and the cell $j$ at time $t+T$, where each time step is represented by a window of 16 days.
The weight of each edge represents the number of drifters following that path.

\begin{figure}[ht]
  \centering
  \includegraphics[width=\linewidth]{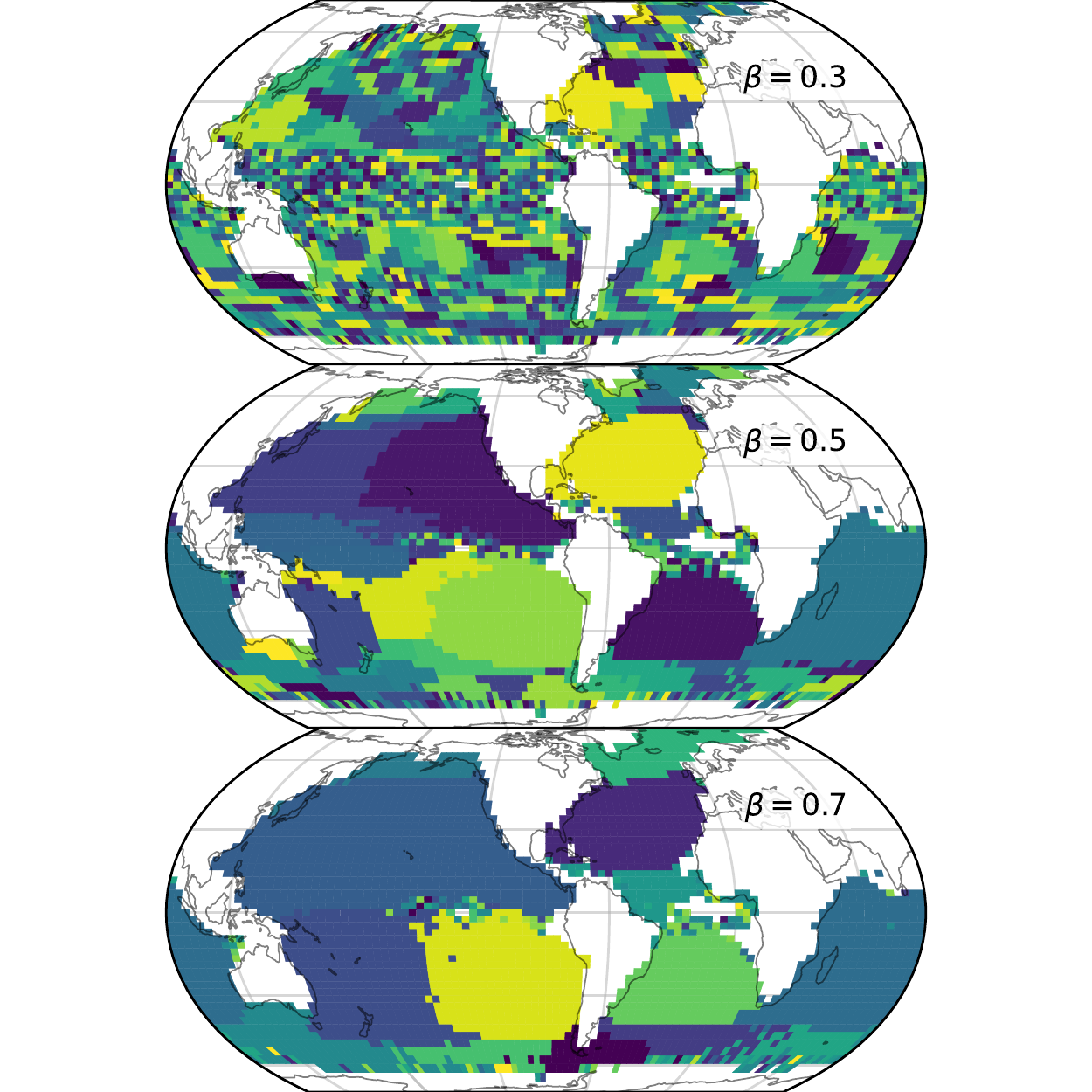}
  \caption{%
    Partitioning the oceans according to the found aggregation classes of the ocean drifters for long time-scales.
    For $T=10$ (about 160 days) the partitions encompass large geographically coherent patches of the ocean surfaces and eventually only subdivide the ocean into well known macro areas.
  }\label{fig:drifters-long}
\end{figure}

In Figures~\ref{fig:drifters-short} and~\ref{fig:drifters-long} we show the partition of the surface currents for a number of values of the regularization parameter $\beta$ and temporal parameter $T$.
Note that for $T=1$, even for high values of the regularization parameter $\beta$, the partitions remain small compared to the ocean gyres.
We estimate the time spent by a drifter to cross the ocean  on one of the major currents to be around 160 days (this is an approximation, since drifter velocities and ocean perimeter are heterogeneous).
For longer time scales ($T=10$, i.e.\ around 160 days) the size of the aggregation classes becomes closer to the size of the major attractors, corresponding to the location of the Garbage Patches (Eastern and Western Great Pacific Garbage Patches, Northern Atlantic Garbage Patch, Indian Ocean Garbage Patch).
The cells visited by fewer drifters are hard to classify: We noticed that in many cases singletons and small classes coincidence with cells with very low visiting probability.

Earlier works focusing on the Mediterranean Sea~\cite{berline2014mediterranean_clustering,rossi2014hydrodynamic} or the Great Barrier Reef~\cite{thomas2014numerical} use an approach similar to~\cite{sonnewald2019ocean_Kmeans}, and use clustering or community detection to identify certain regions of the oceans based on a simulated dynamical model.
Our partition also shows strong similarities with empirical data of the phase in the annual oscillation of the elevation of the surface of the ocean (Figure 7b in~\cite{wunsch1998satellite}), another proxy for the dynamical behavior.
Other approaches include measuring and thresholding dynamical similarities from a set of climatological or oceanographic measures~\cite{donges2009complex, molkenthin2014network_from_flows, tupikina2016correlation_net_flows} to construct the topological structure of a network model.

\begin{figure}[ht]
  \centering
  \includegraphics[width=\linewidth]{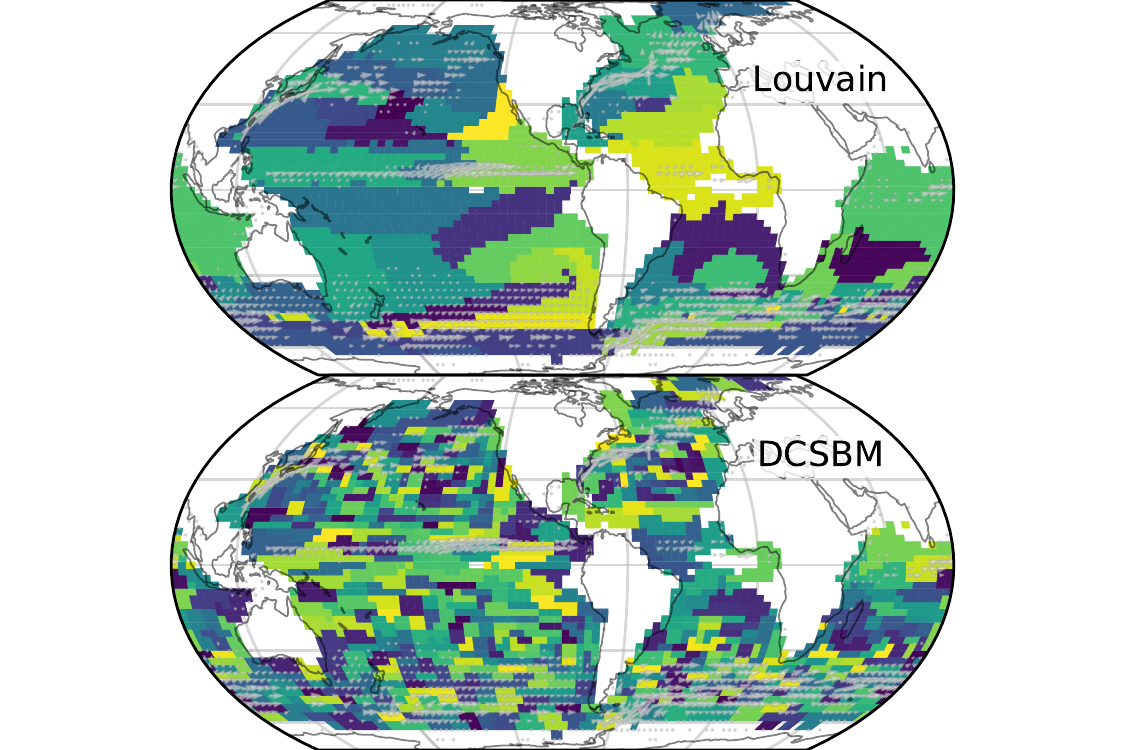}
  \caption{%
    Partitioning the oceans according to other algorithms.
    In particular the outcome of the Louvain algorithm (above) and the DCSBM inference provided by the \texttt{graph-tool} implementation.
    The partitioning of the ocean appears to be overfitted in both cases.
  }\label{fig:drifters-algos}
\end{figure}

\section{Perfect Markovian aggregation: lumpable Markov chains and equitable partitions}

A Markov chain is said to be \emph{lumpable}~\cite{buchholz1994exact,tian2006lumpability} if we can aggregate its states $x_t$ such that the dynamics of the aggregated states $y_t$ is again a Markov chain.
Let us define the partition induced by the classes of the state aggregation by the indicator matrix $Z \in {\{0,1\}}^{N\times K}$ with entries $Z_{xy} = 1$ if state $x$ is mapped to the aggregated state $y= h(x)$ and zero otherwise.

We can then write the condition of a Markov chain to be lumpable in terms of the following algebraic relation:
\begin{equation}\label{eq:lumpable}
PZ = ZP^\pi
\end{equation}
where we have denoted the transition matrix of a Markov chain by $P$, and we have defined the aggregated state transition matrix as $P^\pi = {({Z}^{T}Z)}^{-1}{Z}^T P Z$.
In other words the above equation asserts that if two states $x$ and $x'$ are mapped to the same aggregated state $h(x) = h(x') = y$ (they belong to the same class), then the probability to transition from $x$ or $x'$ to the whole aggregation class $h^{-1}(y')$ defined by the aggregated state $y'$ will only depend on $y$ and $y'$, but not on $x$ or $x'$.
The condition in Eq.~\eqref{eq:lumpable} implies that when a transition $y \to y'$ is observed in the aggregated process, knowing the exact state $x$ inside the aggregation class $h^{-1}(y)$ observed in the original process is of no help to predict the future of the process, since all other states in the same aggregation class lead to the same statistics for the future trajectory.

Note that if a Markov chain is lumpable we can effectively use the smaller transition matrix $P^{\pi}$ to simulate the full chain exactly within the projected subspace.
Hence, if we can find a lumpable partition, we can significantly simplify the description of the system dynamics.
Accordingly, finding such lumpable partitions is of high-interest from a dynamical perspective.
There are a number of important consequences of the above algebraic relationship (cf.~\cite{Schaub2016,o2013observability}): (i) it implies that the transition matrix $P$ will have a set of eigenvectors, such that each eigenvectors is piecewise constant on every aggregation class; (ii) these eigenvectors correspond to appropriately rescaled eigenvectors of the aggregated transition matrix $P^{\pi}$; (iii) the eigenvalues associated to these (scaled) eigenvectors are the same for both $P$ and $P^\pi$, i.e., the eigenvalues of $P^\pi$ are a subset of the eigenvalues of $P$.
Importantly, however, those eigenvalues do not have to correspond to dominant modes of $P$.

Interestingly, the algebraic condition for a Markov chain to be lumpable is closely related to so-called equitable partitions of (directed or undirected) graphs~\cite{Schaub2016,godsil2013algebraic}.
An equitable partition, splits the graph into classes of nodes $\{\mathcal C_i\}$ such that the number of connections from any node $v\in\mathcal C_i$ to a class $\mathcal C_j$ is only dependent on $\mathcal C_i$ and  $\mathcal C_j$, but not on $v$.
Similar to above, let us define the indicator matrix $\mathcal Z$ with entries $\mathcal Z_{ij}=1$ if node $i$ is in class $j$.
Then the algebraic characterization of an equitable partition reads as follows:
\begin{equation}
A\mathcal Z = \mathcal ZA^\pi,
\end{equation}
where $A^\pi = (\mathcal Z^{T}\mathcal Z)^{-1}\mathcal Z^T A \mathcal Z$ can be interpreted as the adjacency matrix of the quotient graph associated to partition $\mathcal Z$.
The quotient graph is defined as follows: the nodes are the aggregation classes $\{\mathcal C_i\}$, and the number of edges between $\mathcal C_i$ and $\mathcal C_j$ is the number of edges in the graph from each node of  $\mathcal C_i$ to each node of class $\mathcal C_j$.
As we can see, the condition for a lumpable Markov chain is identical to that of an equitable partition, apart from the fact that in one case we use the transition matrix $P$ and in the other the adjacency matrix $A$.
This implies (as can be verified by direct computation) that if we perform an unbiased random walk on the graph, then any equitable partition is lumpable for the random walk process.
Note that the equitability is in fact a stronger requirement: equitability regards the absolute number of links going from a node to any class, while lumpability for the random walk is formulated in terms of  the (relative) fraction of links going from a node to any aggregation class.
In the following we will use the above relationship between lumpable Markov chains and equitable partitions to construct lumpable Markov chains.

\begin{figure}[htpb]
	\centering
	\includegraphics[width=\linewidth]{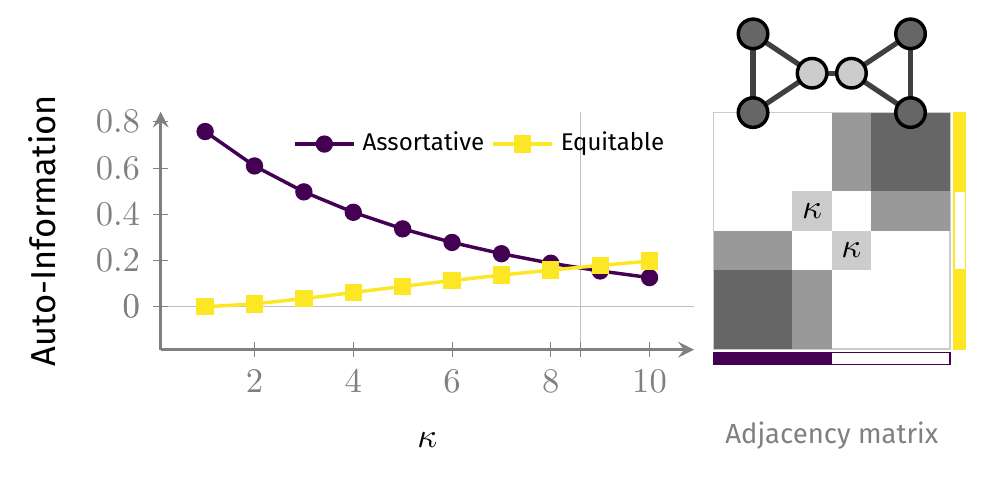}
	\caption{%
		A family of graphs with two dynamically meaningful partitions (assortative and equitable) inspired by the small depicted bow-tie graph.
		Most of the tested algorithms select the assortative partition as the best description for such system.
		By tuning the connectivity parameter $\kappa$ the separation of timescales in the dynamics can be controlled.
		Depending on the mutual relation of the timescales of the dynamics on assortative and equitable partitions, either one or the other can display higher autoinformation.
		(Numerical values computed with $T=1$ and $\beta=0$).
	}\label{fig:bowtie}
\end{figure}

\section{Detecting equitable partitions via autoinformation}
In Fig.~\ref{fig:bowtie} we show a family of graphs with two dynamically meaningful partitions: an assortative partition with low number of edges between aggregated classes; and an equitable partition.

At low values of the number of links between the assortative classes ($\kappa$), the assortative partition has higher autoinformation due to the high predictability of the projected dynamics together with its \emph{almost} Markovianity.
Accordingly, a spectral analysis of the graph Laplacian would split the nodes in the same assortative partition.
In fact, using GenPCCA, fitting a DCSBM, and maximizing modularity, all prefer the assortative partition associated to the highest eigenvalue of the system if we specify that partition in two classes is to be found, regardless of the value of $\kappa$.

However, projecting Markovian dynamics to an equitable partition leads to exactly Markovian dynamics on the class space.
When we increase $\kappa$ the characteristic timescales of the dynamics on the assortative and equitable partitions become more similar but the projected dynamics associated to the assortative partition get \emph{less} Markovian.
By construction of the autoinformation, the fact that an equitable partition induces a Markov dynamics eventually leads to a higher autoinformation for large $\kappa$ in such partition.

Interestingly, we never observe an eigenvalue crossing for the values of $\kappa$ considered here, i.e., the dominant (slowest) modes in terms of eigenvalues are always those associated to the assortative split.
This split, however, induces a non-Markovian projected dynamics.
Thus if we were to purely concentrate on \emph{dominant (i.e.\ slow) timescales, rather than Markovianity, the assortative partition would always be favoured}--- this is exactly the case for spectral algorithms such as GenPCCA that focus on dominant eigenvectors and are thus unable to detect in this case the equitable structure.

\begin{figure}[htpb]
	\centering
	\includegraphics[width=\linewidth]{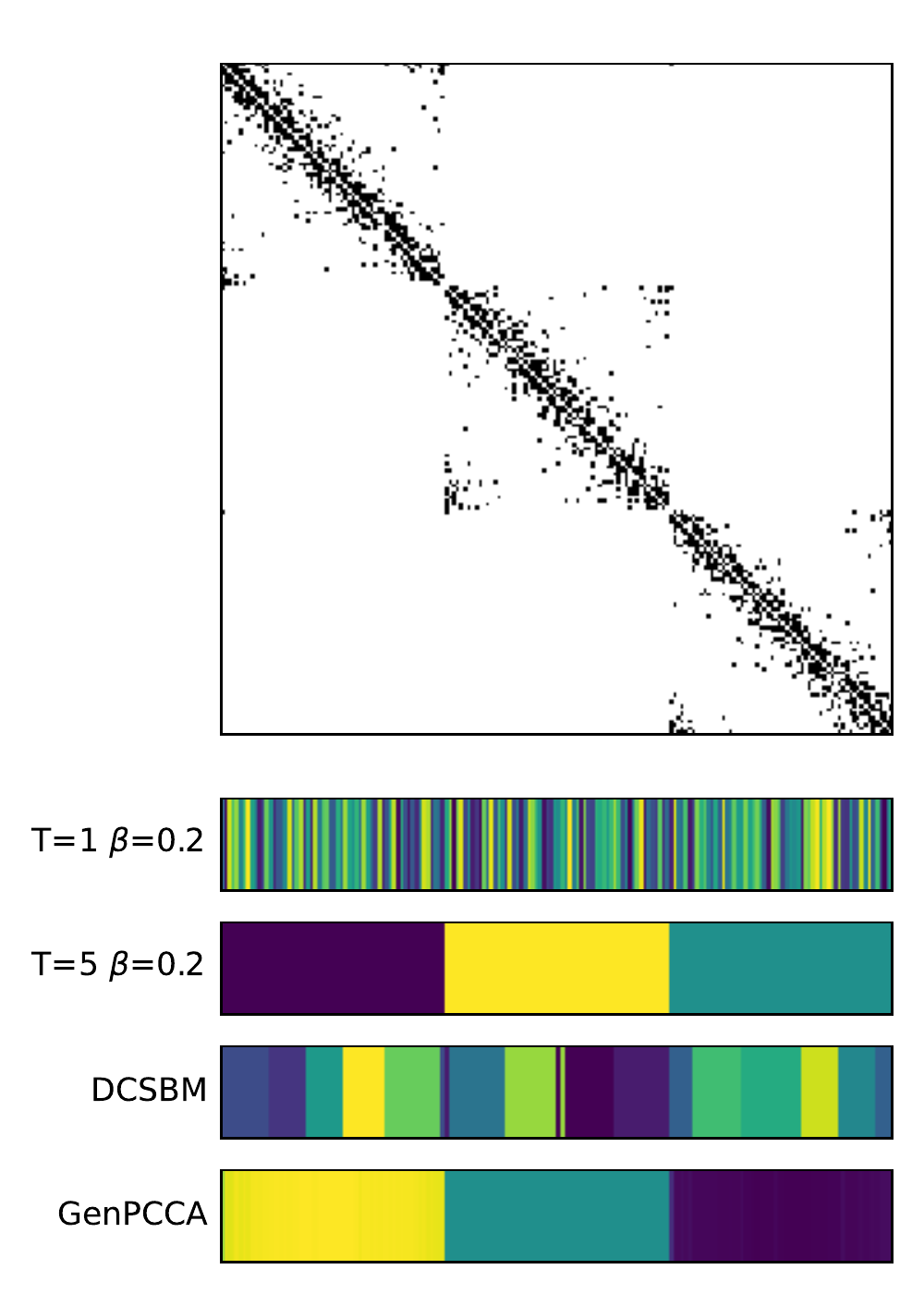}
	\caption{%
		Range dependant network with non-negligible connectivity probability within classes.
		We apply statistical inference of a DCSBM, GenPCCA as well as autoinformation maximization to the graph described by the adjacency matrix on the top.
		The statistical inference of a DCSBM overfits the graph with a high number of small classes along the building cycles.
		GenPCCA and autoinformation maximization recover the planted partition (if we choose a temporal or model selection parameter higher enough).
		Building parameters are: $N=180$, $\alpha = 0.9$, $\gamma=0.8$.
	}%
	\label{fig:stoch}
\end{figure}

\begin{figure}[htpb]
	\centering
	\includegraphics[width=\linewidth]{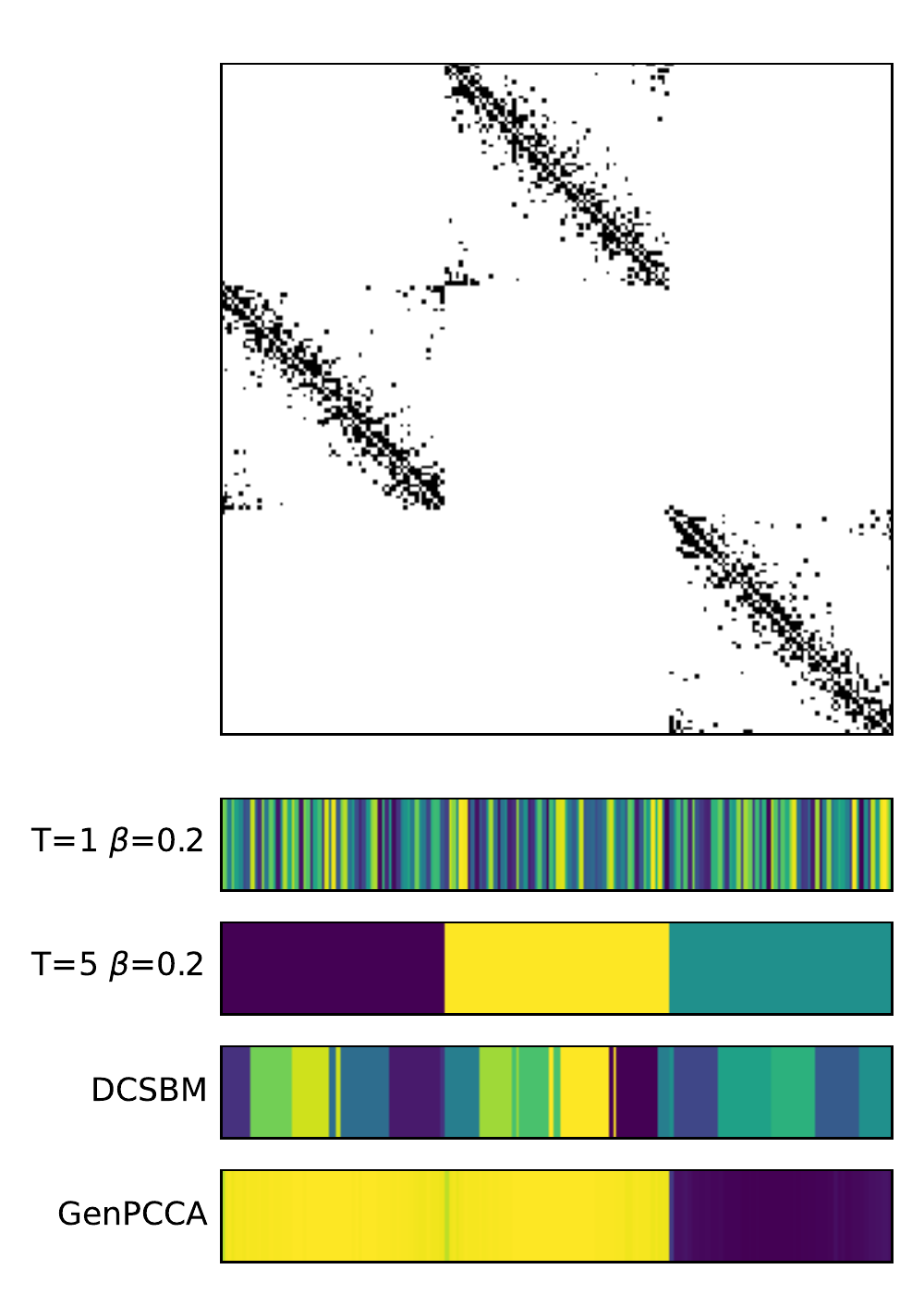}
	\caption{%
		Range dependant network with non-negligible connectivity probability within and between classes (same parameters as in Fig.~\ref{fig:stoch}).
		We apply statistical inference of a DCSBM, GenPCCA as well as autoinformation maximization to the graph described by the adjacency matrix on the top, the first two classes have links mostly between them.
		The statistical inference of a DCSBM overfits the graph with a high number of small classes along the building cycles.
		In this case GenPCCA underfits the original partition recognising only the structure highlighted by the spectral analysis.
		Autoinformation maximization recovers the planted partition (if we choose a temporal parameter $T$ higher enough).
	}%
	\label{fig:stoch90}
\end{figure}

\section{Comparisons with other partitioning algorithms}

The previous section allows to compare on a toy example a spectral method and autoinformation maximization. In this section we further compare partitioning methods on synthetic and real-life examples.

The network in Fig.~\ref{fig:sameB} in the main text represents the state-transition graph of a Markov chain, for which two dynamically relevant partitions can be defined: an assortative partition consisting of the two ring-like structures, and an almost bipartite partition.
Note that the graph can be represented by a conjunction of two \emph{banded adjacency matrices}.
For any banded adjacency matrix, we can be reorder the nodes such that most non-zero entries are close to the main diagonal in a staircase like pattern.

Importantly, for graphs described by banded adjacency matrices most nodes are not directly linked, but are only connected through a chain of connections (a ring like backbone of the graph).
Accordingly, algorithms relying only on information encoded in paths of length one (edges), are likely to be unable to capture this structure.
In this specific case shown in Fig.~\ref{fig:sameB}, using statistical inference to infer a DCSBM will lead to an over-partitioning of the network into many classes.
For instance, the algorithm described in~\cite{peixotoPRL2013parsimonious} finds 18 classes.
We emphasize this is a generic result that can be observed with other techniques as well: Modularity optimization, e.g., via the Louvain algorithm finds 7 classes.
Likewise, statistical inference based on the DCSBM as provided by the  implementation~\cite{peixotoPRL2013parsimonious} as well as Modularity maximization via the Louvain algorithm will partition the ocean surface with a high number of classes  (see Fig.~\ref{fig:drifters-algos}).

Spectral algorithms such as GenPCCA~\cite{fackeldey2018spectral} that aim to identify dominant subspaces of a (transition) matrix, i.e., focus on slow time-scales, can typically resolve banded adjacency structures, as these often induce slow dynamical modes.
Specifically, in the case of Fig.~\ref{fig:sameB} GenPCCA indeed finds the split into the two cyclic structures.
For the ocean example, the runtime of the GenPCCA algorithm was however too long to be compared, so we cannot report results here.

Optimizing regularized autoinformation can be seen as a way to interpolate between structures describing different dynamical modes with possibly different time-scales.
Indeed the autoinformation identifies both dynamically meaningful partitions for a corresponding time-scale parameter $T$ in Fig.~\ref{fig:sameB}.

\subsection{Range dependent networks}
\label{ssec:rangedependent}

Here, we introduce graph ensembles for which similar effects than in~\ref{fig:sameB} in the main text can be observed, when applying graph partitioning algorithms.
The construction of these graphs is closely related to the so-called range-dependent graphs~\cite{grindrod2002rangedependent}, and the $\mathbb{S}_1$ model~\cite{Serrano2008}.

Specifically, we consider networks composed of $N$ nodes endowed with a class label $c_i$, just like in a stochastic block model.
In addition, in each class every node is equipped with an angular coordinate $\theta_i$, corresponding to a point on the unit circle.
We denote by $d_\theta(i,j)$ the (shorter) angular distance between the two coordinates on the (possibly different) circles.
For our construction below we assume that, within each class the nodes have angular coordinates uniformly spaced on the circle.

We define a distance between two nodes $i$ and $j$ as:
\begin{equation*}
  d_{ij} = \frac{d_\theta(i, j) \cdot \sqrt{N_{c_i} N_{c_j}}}{2\pi}
\end{equation*}
where $N_c$ is the size of class $c$.
We now connect two nodes $i$ and $j$ with a probability $p_{ij}=f(d_{ij},c_i,c_j)$ that depends on the above defined distance along the circle between the two nodes and the class labels:
\begin{equation*}
    p_{ij} = \alpha_{c_i,c_j} \cdot (\gamma_{c_i,c_j})^{d_{ij}}
\end{equation*}
where $\alpha_{c_i,c_j}\in[0,1]$ defines a purely class specific connectivity, and $\gamma_{c_i,c_j}\in[0,1]$ is a class dependent parameter that modulates the influence of the distance.
We set $p_{ii}=0$ to ensure that there are no self-loops.
Note that if $\gamma_{c_i,c_j}=1$ for all class labels, we recover the SBM where the probability to link between nodes of two classes is simply given by $\alpha_{c_i,c_j}$.
For $\gamma_{c_i,c_j}\neq 1$, the link probability depends on the angular distance between the two nodes: the larger the distance, the smaller the connection probability.
Hence if $\gamma_{c_i,c_i} <1$ the link probability within each $c_i$ block will be akin to a stochastic cycle.
If $\gamma_{c_i,c_j} <1$ with $i \neq j$, the link probability will be high between corresponding nodes of the two stochastic cycles.
For simplicity we consider the distance parameter to assume a fixed value $\gamma_{c_i, c_j} = \gamma$.

We now consider two types of networks generated according to the above outlined constructions, as exemplified by the adjacency matrices in Figures~\ref{fig:stoch} and \ref{fig:stoch90}.
In the first scenario (Fig.~\ref{fig:stoch}) we consider an assortative setup in which the class-dependent connectivity $\alpha_{c_i, c_k} = \alpha$ and the range parameter $\gamma_{c_i, c_k}=\gamma=0.8$ if $c_i = c_j$, and very small ($\alpha_{c_i, c_j}=\varepsilon \ll 1$ with $\gamma=1$) otherwise.
The resulting adjacency matrix thus consists of a set of banded matrices along the diagonal blocks (see Fig.~\ref{fig:stoch}, top).

For a typical graph drawn from this ensemble Fig.~\ref{fig:stoch} also shows a comparison of the partitioning results obtained from the autoinformation maximisation, the statistical inference of a DCSBM, and partitioning with GenPCCA.
The statistical inference of a DCSBM appears to overfit the graph and tends to find many small blocks within each of the stochastic cycles, on the other hand spectral approaches as well as our dynamical based approach can detect the planted partition of three classes.

In Fig.~\ref{fig:stoch90}, we consider a similar scenario, where $\alpha_{c_i, c_j} = \varepsilon$ apart from the specific interclass link probability $\alpha_{1,2}=\alpha_{2,1}$ and the within class link probability $\alpha_{3,3}$.
For the statistical inference of the DCSBM we find again similar results as before.
However, in this case GenPCCA lumps together the first two planted classes and cannot resolve both the assortative and disassortative cyclic blocks, which is in agreement with the shape of the dominant eigenvectors.
In contrast, the maximization of autoinformation resolves the planted partition structure correctly and unveils both assortative and disassortative features.

\section{How many aggregation classes? Practical recommendations}

In many unsupervised data minign methods, one is confronted to a trade-off between the complexity of a representation of the original data and its accuracy in reproducing some features of the data.

In clustering or partition methods, this amounts to choosing the number of clusters or communities or (in our case) aggregation classes $K$.

In some applications, the number of classes $K$ is known to the user and may be directly imposed to the algorithm.
In this case one seeks the partition of the Markov chain states into $K$ classes that results into the highest autoinformation of the resulting aggregated process.

In other cases however, the number of classes is not known to the user, or only approximately so. In this case, various methods for selecting $K$ are used, some tailored for a particular algorithm, some relatively generic as they can be adapted to a wide diversity of partitioning algorithms.

Here we discuss two generic strategies for choosing $K$, which we recommend for their simplicity and versatility. The reader is of course free to combine our autoinformation criterion with any other heuristics of their preference.

The first generic heuristics is the \emph{elbow criterion}~\cite{kodinariya2013review}, which in this case works as follows:

\begin{enumerate}
	\item Choose an interval of interest for $K$ (which can be as large as $[1,N]$).
    \item For each $K$ in this interval, find the partition into $K$ aggregation classes with highest autoinformation, with the heuristics in \ref{alg} (with $c_{\min} = c_{\max} = K$ and $\beta=0$).
	\item Plot the highest autoinformation found for $K$ classes as an (increasing) function of $K$.
	\item Look for elbows in the plot, i.e. values of $K$ that mark a break from a steady increase of autoinformation (for  $\leq K$ classes) to a slower increase (for $\geq K$ classes).
    \item Each elbow value for $K$ is deemed to represent a \emph{natural} value for $K$, beyond which the \emph{quality} of the reduced model increases at slower pace.
\end{enumerate}

This simple and intuitive method is effective in some circumstances but may prove too crude in others, and leave no clear conclusions.
One of the reasons is that $K$ alone is not always a good representation of the \emph{complexity} or \emph{size} of the aggregated model.
For example a split of 100 states into three classes of 50, 45 and 5 nodes may be interpreted as \emph{essentially} a split into two classes with a small ``correction'' in terms of a small third class.
A metric that takes into account such heterogeneity of classes is the entropy $H$ of the partition.
Optimising the autoinformation with a constraint on the maximal allowed partition entropy $H$ (in lieu of $K$) is equivalent to the regularised autoinformation, with $\beta$ as a Lagrange multiplier.
The choice of the number of classes is now replaced with the choice of a parameter $\beta$, controlling the trade-off between high autoinformation and low entropy of the aggregated model.
In this framework, a heuristic recommended is thus the \emph{plateau} or \emph{robustness} criterion~\cite{lambiotte2010multi,delmotte2011protein}, which checks for a plateau, i.e. in our case a large interval of $\beta$ where the solution is robust, in that it keeps the same number of clusters:

\begin{enumerate}
  \item Choose a set $B=\{\beta_i\}$ of values of interest for the regularization parameter $\beta$ (which can be a discrete sampling of $[0,1]$).
  \item For each value $\beta_i \in B$, find the partition with highest regularised autoinformation, with the heuristics in \ref{alg} (setting $c_{\min} = 1, c_{\max} = N$ and $\beta=\beta_i$).
  \item Plot the number of classes $K$ found as a (decreasing) function of $\beta$. Typically this plot present plateaux.
  \item Look for plateaux in the plot, i.e. interval of values of $\beta$ for which the algorithm finds the same partition, thus with a constant number of classes.
  \item Each large plateau is deemed to represent a \emph{natural partition} of the graph, robust to the choice of $\beta$.
\end{enumerate}

Ultimately the problem of selecting $K$ is ill-defined, as different users faced to different applications may wish different trade-offs between quality of the description and complexity of the model.
These, and more advanced methodologies, are simply tools to help the user make an informed decision.
Fully automated model selection procedures also exist in some cases (e.g. modularity maximisation or Bayesian DCSBM inference), that implicitly internalise a certain choice strategy for the user.

We now illustrate the elbow criterion and the plateau criterion on an example where a \emph{ground truth} structure is planted, and we show that both criteria are able to recover the planted structure.

\begin{figure}[htpb]
  \centering
  \includegraphics[width=0.8\linewidth]{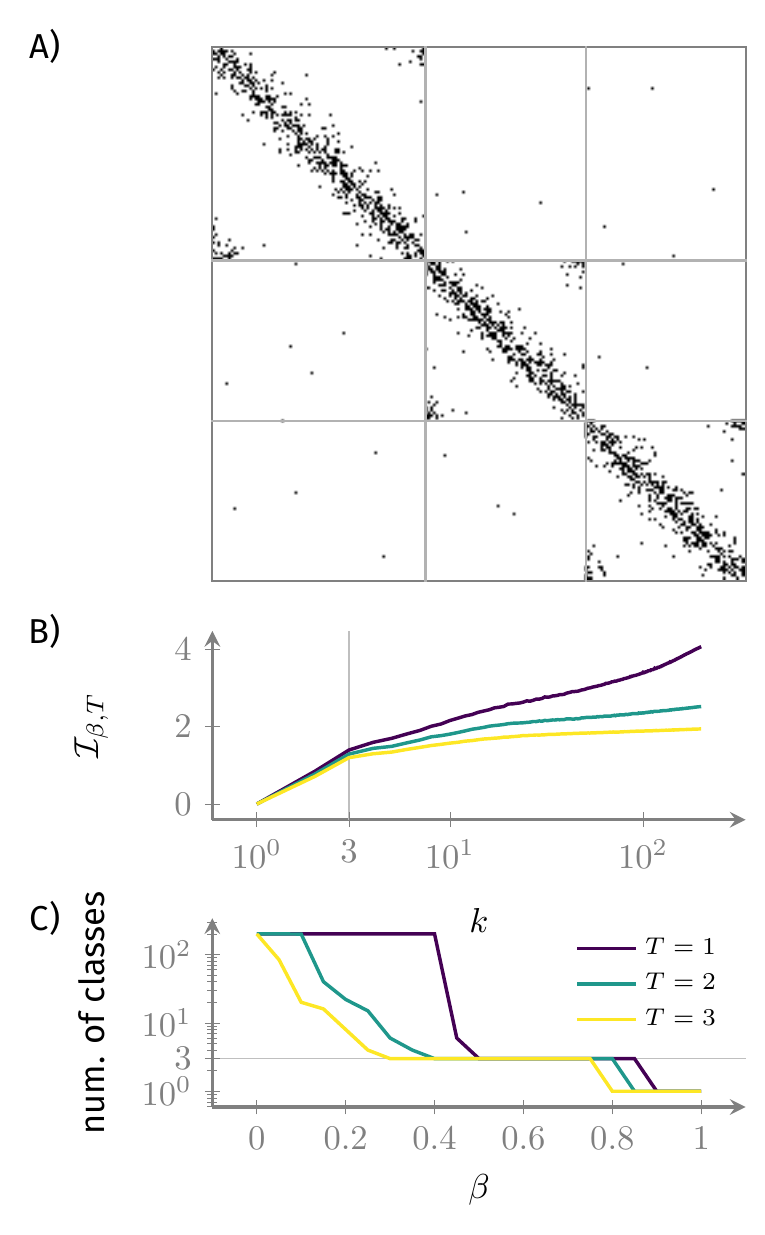}
  \caption{Model selection on range dependent networks.
    Maximization of autoinformation is performed on the range dependent network composed by three planted partitions and with adjacency matrix depicted in (A).
    Autoinformation maximized at fixed number of classes $k$ (B) shows an elbow when the number of classes corresponds to the planted partition.
    Maximization of autoinformation at different values of $\beta$ (C) shows a plateau at the number of classes of the planted partition.
    The graph is built with three classes of 80, 60 and 60 nodes respectively.
    Diagonal blocks have $\gamma=0.8$ and $\alpha=0.9$; out of diagonal blocks have $\gamma=1$ and $\alpha=0.001$.
  }%
  \label{fig:modelselection}
\end{figure}

In Fig.~\ref{fig:modelselection} we maximize autoinformation in a range dependent network (see above for a description of the generative model).
The network is built with 200 nodes divided into one class of 80 nodes and two classes of 60.
We set $\gamma_{c_i, c_j}=\gamma=0.8$ and $\alpha_{c_i, c_j}=\alpha=0.9$ within classes ($c_i=c_j$), while $\gamma=1$ and $\alpha=\varepsilon \ll 1$ between classes.

Maximization of autoinformation for different fixed numbers of $k$ of classes (see Fig.~\ref{fig:modelselection}B) shows an elbow in the value of autoinformation.
The elbow corresponds to the number of classes in the planted partition and becomes more prominent as $T$ increases.
Fitting a DCSBM with the fully automated criterion proposed in~\cite{peixotoPRL2013parsimonious} to the same graph, results in a partition with 9 to 11 classes (consistently with our previous examples).

The results of using the complexity parameter $\beta$ to select the number of classes are shown in Fig.~\ref{fig:modelselection}C.
In this case a wide plateau is found when the partition that maximizes autoinformation has the planted number of classes, revealing a partitioning that is robust with respect to the choice of $\beta$.
Increasing the value of $T$ corresponds to a growth of the robustness plateaux and a shift to lower values of $\beta$.

\section{Algorithm}

We briefly describe in Algorithm~\ref{alg} the AutoInformation State-Aggregation algorithm (AISA) used to maximize the autoinformation within this paper.
The algorithm uses an approach inspired by the simulated annealing technique to sample the state space of all possible partitions.
A code with a reference implementation in \texttt{Python} is available as a \texttt{git} repository at the following hyperlink: \href{https://maurofaccin.github.io/aisa}{AISA (https://maurofaccin.github.io/aisa)}.

\begin{algorithm}[ht!]
\SetAlgoLined
\KwResult{Partition of the nodes}
 Optional initialization with initial partition\;
 $T \leftarrow$ time-scale parameter\\
 $\beta \leftarrow$ regularization parameter (default is $0$)\\
 $\mathcal T \leftarrow$ initial pseudo-temperature\\
 $c_{\min}, c_{\max} \leftarrow$ class number bound (default is $[1,N]$)\\
 \While{no accepted move for a max duration or max steps reached}{
  $n_i \leftarrow \textrm{random node}$\\
  $c \leftarrow \textrm{class of node $n_i$}$\\
  $\hat c \leftarrow \textrm{random class}$\\
  $\delta \leftarrow \mathcal I_{\beta, T}(n_i \in \hat c) - \mathcal I_{\beta, T}(n_i \in c)$

  \eIf{$\delta > 0$}{
    node $i$ moved to class $\hat c$
   }{
   $p_{\hat c \rightarrow c} \leftarrow$ probability of moving $n_i$ from $\hat c$ to $c$\\
   $p_{c \rightarrow \hat c} \leftarrow$ probability of moving $n_i$ from $c$ to $\hat c$\\
   $\textrm{prob} \leftarrow p_{\hat c \rightarrow c}/p_{c \rightarrow \hat c}$\\
   $t \leftarrow e^\frac{\delta}{\mathcal T\cdot\textrm{prob}}$\\
   $r \leftarrow \textrm{random number} \in [0, 1]$\\
   \eIf{$r < t$}{
    node $i$ moved to class $\hat c$
  }{\textrm{nothing is done}}
  }
  $\mathcal T$ is decreased
 }
 \caption{Pseudocode used to optimize autoinformation}
 \label{alg}
\end{algorithm}

With the algorithm in \ref{alg}, one can maximize the autoinformation fixing the number of classes.
This is achieved by fixing the value of $\beta$ (defaults to $0$) and assigning the same value to the class number bounds ($c_{\min} = c_{\max} =k$).
In this way the moves that change the total number of classes are forbidden.

Similarly one may prefer to set the regularization parameter $\beta$ to some value and let the algorithm maximize the regularized autoinformation over a wide interval of values of $k$.

\section{Related literature}\label{connection_to_previous_works}
In this section, we comment on similarities and differences to previous works that appeared in the literature, complementing the discussion on limiting cases above.

A number of information theoretic methods have been proposed for the analysis and compression of dynamical data generated by Markov processes.
Computational mechanics~\cite{Crutchfield1989,shalizi2001computational,CrutchfieldFeldman2003,kelly2012new} provides a framework to construct a minimal dynamical description of an observed stationary process in terms of an $\epsilon$-machine, which is a minimal system description commensurate with an accurate description of the process.
This focus on predictability is similar to the approach presented here.
However, our goal is not to find an optimal state space representation of an arbitrary process in terms of predictability.
Instead, we are interested in the opposite direction, we start with a given Markov chain (and its state-space representation) and want to find an approximate description (a ``lossy compression'') of the dynamics.

The information bottleneck method~\cite{Tishby2000} provides another information theoretic method that provides a way to find a compressed (or quantized) representation of a random signal via a variational problem.
In contrast to our method here the information bottleneck method was however not designed with a random process in mind and involves choosing a relevance variable that captures the features one wants the compressed description to preserve.

As the states in Markov process can be interpreted as nodes in a graph, with state transitions encoded by edges, any Markov process can also be mapped to a network and vice versa.
Accordingly such methods have also been employed in the context of the analysis of complex networks~\cite{masuda2017random}.
The map equation framework~\cite{Rosvall29012008} by Rosvall et al.\ proposes to compress the one-step transition properties of random walks on networks under a specific coding scheme.
Similar to our work, finding the optimal compression in terms of the assignment of nodes to codewords within the map equation framework, is also associated to finding a partition of the nodes. However, the coding scheme used in the map equation effectively amounts to a mean-field description of the transition properties of the associated Markov chain~\cite{Schaub2012}.
As a consequence, only densely connected groups of nodes (assortative community structure) are identified by this scheme.

Another approach related our work is the study by Peixoto and Rosvall~\cite{Peixoto2017} which proposes a model for ``Markov chains with community structure'' and uses a Bayesian framework to fit the model.
While~\cite{Peixoto2017} also recovers an DCSBM as a special case of their method their approach \emph{imposes} that the observed dynamics have a transition matrix of a block-diagonal form, i.e.\ their generative model posits \emph{a priori} a certain transition structure.
In contrast, we do not make any \emph{a priori} assumption on the transition matrix of the observed Markov chain, but show that the autoinformation on symmetric networks for $T=1$ is equivalent up trivial transformations to the log-likelihood the DCSBM\@.
Stated differently, both approaches lead under the specific assumption of a reversible dynamics and a time horizon of $T=1$ to equivalent optimization problems --- hence for this special case, we can give alternative interpretations of the corresponding methods (cf.\ Fig.~\ref{fig:sameB}).

Finally, we remark that our framework exhibits certain parallels to the so-called Markov stability framework.
Just like Modularity~\cite{newman2004finding} can be dynamically interpreted as a sum of covariances between successive dynamical states of a random walker~\cite{Delvenne20072010,Schaub2014,Schaub2019}, here we have shown how the use of information theoretic measures can provide us with a dynamic interpretation of the DCSBM\@.
Interestingly, it has been shown by Newman recently~\cite{Newman2016} that the objective function of Modularity can, with a specifically chosen resolution parameter, be also be interpreted as the likelihood of a planted partition model, a particular type of assortative block model.
In contrast, the equivalence we showed here between the autoinformation for $T=1$ and the likelihood function of the DCSBM holds for general block models and is not limited to any specific structure.
See also the discussion on the short and long scale limits above.
\end{document}